\documentclass[longauth]{aa}

\usepackage[utf8]{inputenc}
\usepackage{graphicx}
\usepackage{txfonts}
\usepackage{euclid}
\usepackage{hyperref}
\usepackage[dvipsnames]{xcolor} 
\usepackage[normalem]{ulem}

\def\RIZ{$VIS$}
\def\biaszmean{$\mu_{\Delta z}$}
\def\scatterzmean{$\sigma_{\Delta z}$}
\def\biaszmeanB{\mu_{\Delta z}}

\def\lephare{\texttt{LePhare}}
\def\like{${\mathcal L }(z)$}
\def\Msol{{\rm M}_\odot}
\def\avz{\langle z \rangle}

\def\Euclid{\textit{Euclid}}
\def\photoz{photo-\textit{z}}
\def\specz{spec-\textit{z}}
\def\zPDF{\textit{z}PDF}
\def\obs{\vec{o}}

\usepackage{calrsfs}

\DeclareMathAlphabet{\pazocal}{OMS}{zplm}{m}{n}

\definecolor{lblue}{rgb}{0.1,0.7,1.}
\definecolor{Orange}{rgb}{1.0,0.05,0.15}
\definecolor{Green}{rgb}{0.15,0.45,0.25}
\definecolor{Blue}{rgb}{0.0,0.08,0.65}
\definecolor{Brown}{rgb}{0.7,0.25,0.0}
\definecolor{Pink}{rgb}{1.0,0.05,0.5}

\begin{document} 

\title{\Euclid{} preparation: XI. Mean redshift determination from galaxy redshift probabilities for cosmic shear tomography} \titlerunning{Determination of the mean redshift of tomographic bins}

\author{Euclid Collaboration: O.~Ilbert$^{1}$\thanks{\email{olivier.ilbert@lam.fr}}, S.~de la Torre$^{1}$, N.~Martinet$^{1}$, A.H.~Wright$^{2}$, S.~Paltani$^{3}$, C.~Laigle$^{4}$, I.~Davidzon$^{5}$, E.~Jullo$^{1}$, H.~Hildebrandt$^{2}$, D.C.~Masters$^{6}$
% ...
}

\institute{$^{1}$ Aix-Marseille Univ, CNRS, CNES, LAM, Marseille, France\\
$^{2}$ Ruhr-Universit\"at Bochum, Astronomisches Institut, German Centre for Cosmological Lensing, Universit\"atsstr. 150, 44801 Bochum, Germany\\
$^{3}$ Department of Astronomy, University of Geneva, ch. d'\'Ecogia 16, CH-1290 Versoix, Switzerland\\
$^{4}$ Institut d'Astrophysique de Paris, 98bis Boulevard Arago, F-75014, Paris, France\\
$^{5}$ Cosmic Dawn Center (DAWN), Niels Bohr Institute, University of Copenhagen, Vibenshuset, Lyngbyvej 2, DK-2100 Copenhagen, Denmark\\
$^{6}$ Infrared Processing and Analysis Center, California Institute of Technology, Pasadena, CA 91125, USA\\
$^{7}$ Institute of Cosmology and Gravitation, University of Portsmouth, Portsmouth PO1 3FX, UK\\
$^{8}$ Jodrell Bank Centre for Astrophysics, School of Physics and Astronomy, University of Manchester, Oxford Road, Manchester M13 9PL, UK\\
$^{9}$ INAF-Osservatorio Astronomico di Brera, Via Brera 28, I-20122 Milano, Italy\\
$^{10}$ INAF-Osservatorio di Astrofisica e Scienza dello Spazio di Bologna, Via Piero Gobetti 93/3, I-40129 Bologna, Italy\\
$^{11}$ Mullard Space Science Laboratory, University College London, Holmbury St Mary, Dorking, Surrey RH5 6NT, UK\\
$^{12}$ IFPU, Institute for Fundamental Physics of the Universe, via Beirut 2, 34151 Trieste, Italy\\
$^{13}$ SISSA, International School for Advanced Studies, Via Bonomea 265, I-34136 Trieste TS, Italy\\
$^{14}$ INFN, Sezione di Trieste, Via Valerio 2, I-34127 Trieste TS, Italy\\
$^{15}$ INAF-Osservatorio Astronomico di Trieste, Via G. B. Tiepolo 11, I-34131 Trieste, Italy\\
$^{16}$ Universidad de la Laguna, E-38206, San Crist\'{o}bal de La Laguna, Tenerife, Spain\\
$^{17}$ Instituto de Astrof\'{i}sica de Canarias. Calle V\'{i}a L\`{a}ctea s/n, 38204, San Crist\'{o}bal de la Laguna, Tenerife, Spain\\
$^{18}$ Dipartimento di Fisica e Astronomia, Universit\'a di Bologna, Via Gobetti 93/2, I-40129 Bologna, Italy\\
$^{19}$ INFN-Sezione di Bologna, Viale Berti Pichat 6/2, I-40127 Bologna, Italy\\
$^{20}$ INAF-Osservatorio Astronomico di Padova, Via dell'Osservatorio 5, I-35122 Padova, Italy\\
$^{21}$ Universit\"ats-Sternwarte M\"unchen, Fakult\"at f\"ur Physik, Ludwig-Maximilians-Universit\"at M\"unchen, Scheinerstrasse 1, 81679 M\"unchen, Germany\\
$^{22}$ Max Planck Institute for Extraterrestrial Physics, Giessenbachstr. 1, D-85748 Garching, Germany\\
$^{23}$ INAF-Osservatorio Astrofisico di Torino, Via Osservatorio 20, I-10025 Pino Torinese (TO), Italy\\
$^{24}$ Dipartimento di Fisica - Sezione di Astronomia, Universit\'a di Trieste, Via Tiepolo 11, I-34131 Trieste, Italy\\
$^{25}$ Universit\'e de Paris, CNRS, Astroparticule et Cosmologie, F-75006 Paris, France\\
$^{26}$ INFN-Sezione di Roma Tre, Via della Vasca Navale 84, I-00146, Roma, Italy\\
$^{27}$ Department of Mathematics and Physics, Roma Tre University, Via della Vasca Navale 84, I-00146 Rome, Italy\\
$^{28}$ INAF-Osservatorio Astronomico di Roma, Via Frascati 33, I-00078 Monteporzio Catone, Italy\\
$^{29}$ INAF-Osservatorio Astronomico di Capodimonte, Via Moiariello 16, I-80131 Napoli, Italy\\
$^{30}$ INFN-Bologna, Via Irnerio 46, I-40126 Bologna, Italy\\
$^{31}$ Dipartimento di Fisica e Scienze della Terra, Universit\'a degli Studi di Ferrara, Via Giuseppe Saragat 1, I-44122 Ferrara, Italy\\
$^{32}$ INAF, Istituto di Radioastronomia, Via Piero Gobetti 101, I-40129 Bologna, Italy\\
$^{33}$ Institut de Recherche en Astrophysique et Plan\'etologie (IRAP), Universit\'e de Toulouse, CNRS, UPS, CNES, 14 Av. Edouard Belin, F-31400 Toulouse, France\\
$^{34}$ INFN-Sezione di Torino, Via P. Giuria 1, I-10125 Torino, Italy\\
$^{35}$ Dipartimento di Fisica, Universit\'a degli Studi di Torino, Via P. Giuria 1, I-10125 Torino, Italy\\
$^{36}$ Universit\'e C\^ote d'Azur, Observatoire de la C\^ote d'Azur, CNRS, Laboratoire Lagrange, Bd de l'Observatoire, CS 34229, 06304 Nice cedex 4, France\\
$^{37}$ INAF-IASF Milano, Via Alfonso Corti 12, I-20133 Milano, Italy\\
$^{38}$ Institut de F\'{i}sica d’Altes Energies (IFAE), The Barcelona Institute of Science and Technology, Campus UAB, 08193 Bellaterra (Barcelona), Spain\\
$^{39}$ Instituto de Astrof\'isica e Ci\^encias do Espa\c{c}o, Faculdade de Ci\^encias, Universidade de Lisboa, Tapada da Ajuda, PT-1349-018 Lisboa, Portugal\\
$^{40}$ AIM, CEA, CNRS, Universit\'{e} Paris-Saclay, Universit\'{e} Paris Diderot, Sorbonne Paris Cit\'{e}, F-91191 Gif-sur-Yvette, France\\
$^{41}$ Institute of Space Sciences (ICE, CSIC), Campus UAB, Carrer de Can Magrans, s/n, 08193 Barcelona, Spain\\
$^{42}$ Institut d’Estudis Espacials de Catalunya (IEEC), Carrer Gran Capit\'a 2-4, 08034 Barcelona, Spain\\
$^{43}$ Observatoire de Sauverny, Ecole Polytechnique F\'ed\'erale de Lau- sanne, CH-1290 Versoix, Switzerland\\
$^{44}$ Department of Physics "E. Pancini", University Federico II, Via Cinthia 6, I-80126, Napoli, Italy\\
$^{45}$ INFN section of Naples, Via Cinthia 6, I-80126, Napoli, Italy\\
$^{46}$ INAF-Osservatorio Astrofisico di Arcetri, Largo E. Fermi 5, I-50125, Firenze, Italy\\
$^{47}$ Centre National d'Etudes Spatiales, Toulouse, France\\
$^{48}$ Institute for Astronomy, University of Edinburgh, Royal Observatory, Blackford Hill, Edinburgh EH9 3HJ, UK\\
$^{49}$ European Space Agency/ESRIN, Largo Galileo Galilei 1, 00044 Frascati, Roma, Italy\\
$^{50}$ ESAC/ESA, Camino Bajo del Castillo, s/n., Urb. Villafranca del Castillo, 28692 Villanueva de la Ca\~nada, Madrid, Spain\\
$^{51}$ Univ Lyon, Univ Claude Bernard Lyon 1, CNRS/IN2P3, IP2I Lyon, UMR 5822, F-69622, Villeurbanne, France\\
$^{52}$ Departamento de F\'isica, Faculdade de Ci\^encias, Universidade de Lisboa, Edif\'icio C8, Campo Grande, PT1749-016 Lisboa, Portugal\\
$^{53}$ Instituto de Astrof\'isica e Ci\^encias do Espa\c{c}o, Faculdade de Ci\^encias, Universidade de Lisboa, Campo Grande, PT-1749-016 Lisboa, Portugal\\
$^{54}$ Department of Physics, Oxford University, Keble Road, Oxford OX1 3RH, UK\\
$^{55}$ INFN-Padova, Via Marzolo 8, I-35131 Padova, Italy\\
$^{56}$ University of Lyon, UCB Lyon 1, CNRS/IN2P3, IUF, IP2I Lyon, France\\
$^{57}$ School of Physics, HH Wills Physics Laboratory, University of Bristol, Tyndall Avenue, Bristol, BS8 1TL, UK\\
$^{58}$ Aix-Marseille Univ, CNRS/IN2P3, CPPM, Marseille, France\\
$^{59}$ Department of Physics, P.O. Box 64, 00014 University of Helsinki, Finland\\
$^{60}$ Dipartimento di Fisica "Aldo Pontremoli", Universit\'a degli Studi di Milano, Via Celoria 16, I-20133 Milano, Italy\\
$^{61}$ INFN-Sezione di Milano, Via Celoria 16, I-20133 Milano, Italy\\
$^{62}$ Institute of Theoretical Astrophysics, University of Oslo, P.O. Box 1029 Blindern, N-0315 Oslo, Norway\\
$^{63}$ Jet Propulsion Laboratory, California Institute of Technology, 4800 Oak Grove Drive, Pasadena, CA, 91109, USA\\
$^{64}$ von Hoerner \& Sulger GmbH, Schlo{\ss}Platz 8, D-68723 Schwetzingen, Germany\\
$^{65}$ Max-Planck-Institut f\"ur Astronomie, K\"onigstuhl 17, D-69117 Heidelberg, Germany\\
$^{66}$ Department of Physics and Helsinki Institute of Physics, Gustaf H\"allstr\"omin katu 2, 00014 University of Helsinki, Finland\\
$^{67}$ Universit\'e de Gen\`eve, D\'epartement de Physique Th\'eorique and Centre for Astroparticle Physics, 24 quai Ernest-Ansermet, CH-1211 Gen\`eve 4, Switzerland\\
$^{68}$ NOVA optical infrared instrumentation group at ASTRON, Oude Hoogeveensedijk 4, 7991PD, Dwingeloo, The Netherlands\\
$^{69}$ Argelander-Institut f\"ur Astronomie, Universit\"at Bonn, Auf dem H\"ugel 71, 53121 Bonn, Germany\\
$^{70}$ Institute for Computational Cosmology, Department of Physics, Durham University, South Road, Durham, DH1 3LE, UK\\
$^{71}$ Institut f\"ur Theoretische Physik, University of Heidelberg, Philosophenweg 16, 69120 Heidelberg, Germany\\
$^{72}$ Zentrum f\"ur Astronomie, Universit\"at Heidelberg, Philosophenweg 12, D- 69120 Heidelberg, Germany\\
$^{73}$ INAF-IASF Bologna, Via Piero Gobetti 101, I-40129 Bologna, Italy\\
$^{74}$ Universit\'e de Paris, F-75013, Paris, France, LERMA, Observatoire de Paris, PSL Research University, CNRS, Sorbonne Universit\'e, F-75014 Paris, France\\
$^{75}$ CEA Saclay, DFR/IRFU, Service d'Astrophysique, Bat. 709, 91191 Gif-sur-Yvette, France\\
$^{76}$ IRFU, CEA, Universit\'e Paris-Saclay F-91191 Gif-sur-Yvette Cedex, France\\
$^{77}$ ICC\&CEA, Department of Physics, Durham University, South Road, DH1 3LE, UK\\
$^{78}$ Department of Physics and Astronomy, University of Aarhus, Ny Munkegade 120, DK–8000 Aarhus C, Denmark\\
$^{79}$ Space Science Data Center, Italian Space Agency, via del Politecnico snc, 00133 Roma, Italy\\
$^{80}$ Institute of Space Science, Bucharest, Ro-077125, Romania\\
$^{81}$ Institute for Computational Science, University of Zurich, Winterthurerstrasse 190, 8057 Zurich, Switzerland\\
$^{82}$ Dipartimento di Fisica e Astronomia “G.Galilei", Universit\'a di Padova, Via Marzolo 8, I-35131 Padova, Italy\\
$^{83}$ Departamento de F\'isica, FCFM, Universidad de Chile, Blanco Encalada 2008, Santiago, Chile\\
$^{84}$ Centro de Investigaciones Energ\'eticas, Medioambientales y Tecnol\'ogicas (CIEMAT), Avenida Complutense 40, 28040 Madrid, Spain\\
$^{85}$ Universidad Polit\'ecnica de Cartagena, Departamento de Electr\'onica y Tecnolog\'ia de Computadoras, 30202 Cartagena, Spain\\
$^{86}$ Kapteyn Astronomical Institute, University of Groningen, PO Box 800, 9700 AV Groningen, The Netherlands\\
$^{87}$ Department of Physics, P.O.Box 35 (YFL), 40014 University of Jyv\"askyl\"a, Finland\\
$^{88}$ Department of Physics and Astronomy, University College London, Gower Street, London WC1E 6BT, UK\\
}

\date{Received on date, accepted on date}

\abstract{
The analysis of weak gravitational lensing in wide-field imaging surveys is considered to be a major cosmological probe of dark energy. Our capacity to constrain the dark energy equation of state relies on the accurate knowledge of the galaxy mean redshift $\avz$. We investigate the possibility of measuring $\avz$ with an accuracy better than $0.002\,(1+z)$, in ten tomographic bins spanning the redshift interval $0.2<z<2.2$, the requirements for the cosmic shear analysis of \Euclid{}. We implement a sufficiently realistic simulation to understand the advantages, complementarity, but also shortcoming of two standard  approaches: the direct calibration of $\avz$ with a dedicated spectroscopic sample and the combination of the photometric redshift probability distribution function (\zPDF{}) of individual galaxies. We base our study on the Horizon-AGN hydrodynamical simulation that we analyse with a standard galaxy spectral energy distribution template-fitting code. Such procedure produces photometric redshifts with realistic biases, precision and failure rate. We find that the \Euclid{} current design for direct calibration is sufficiently robust to reach the requirement on the mean redshift, provided that the purity level of the spectroscopic sample is maintained at an extremely high level of $>99.8\%$. The \zPDF{} approach could also be successful if we debias the \zPDF{} using a spectroscopic training sample. This approach requires deep imaging data, but is weakly sensitive to spectroscopic redshift failures in the training sample. We improve the debiasing method and confirm our finding by applying it to real-world weak-lensing data sets (COSMOS and KiDS+VIKING-450).

}


\authorrunning{Euclid Collaboration: O. Ilbert et al.}

\maketitle

\section{Introduction}
\label{sec:introduction}

Understanding the late, accelerated expansion of our Universe \citep{riess98, perlmutter99} is one of the most important challenges in modern cosmology. Three leading hypotheses are: a modification of the laws of gravity, the introduction of a cosmological constant $\Lambda$ in the equations describing the dynamics of our Universe, or the existence of a dark energy fluid with negative pressure. The two latter hypotheses could be disentangled one from another by measuring the equation of state $w$ of dark energy, which links its pressure to its density. Only the case $w=-1$ is compatible with a cosmological constant, and therefore any deviation from this value would invalidate the standard $\Lambda$ cold dark matter ($\Lambda$CDM) model, in favour of dark energy. This makes the precise measurement of $w$ a key component of future cosmological experiments such as \Euclid{} \citep{laureijs11}, the Vera C. Rubin Observatory Legacy Survey of Space and Time \citep[LSST;][]{LSSTsciencebook}, or the \textit{Nancy Grace Roman} Space Telescope \citep{spergel15}.

Cosmic shear \citep[see e.g.][for recent reviews]{kilbinger15,mandelbaum18}, which is the coherent distortion of galaxy images by large-scale structures via weak gravitational lensing, offers the potential to measure $w$ with great precision: the \Euclid{} survey, in particular, aims at reaching $1\%$ precision on the measurement of $w$ using cosmic shear. One advantage of using lensing to measure $w$, compared to other probes, is that there exists a direct link between galaxy image geometrical distortions (i.e. the shear) and the gravitational potential of the intervening structures. When the shapes of, and distances to, galaxy sources are known, gravitational lensing allows one to probe the matter distribution of the Universe. 

This discovery has led to the rapid growth of interest in using cosmic shear as a key cosmological probe, as evidenced by its successful application to several surveys. Constraints on the matter density parameter $\Omega_{\rm m}$, and the normalisation of the linear matter power spectrum $\sigma_8$, have been reported by the Canada-France-Hawaii Telescope Lensing Survey \citep[CFHTLenS,][]{kilbinger13}, the Kilo Degree Survey \citep[KiDS,]{hildebrandt17}, the Dark Energy Survey \citep[DES,][]{troxel18}, and the Hyper-Suprime Camera Survey \citep[HSC,][]{hikage19}. These studies typically utilise so-called cosmic shear tomography \citep{Hu99}, whereby the cosmic shear signal is obtained by measuring the cross-correlation between galaxy shapes in different bins along the line of sight (i.e. tomographic bins). 
Large forthcoming surveys, also utilising cosmic shear tomography, will enhance the precision of cosmological parameter measurements (e.g. $\Omega_{\rm m}$, $\sigma_8$, and $w$), while also enabling the measurement of any evolution in the dark-energy equation of state, such as that parametrised by \citet{caldwell98}: $w=w_0 \, + \, w_a \, (1 - a)$, where $a$ is the scale factor. 

Tomographic cosmic shear studies require accurate knowledge of the galaxy redshift distribution. The estimation and calibration of the redshift distribution has been identified as one of the most problematic tasks in current cosmic shear surveys, as systematic bias in the distribution calibration directly influences the resulting cosmological parameter estimates. In particular, \citet{joudaki20} show that the $\Omega_{\rm m}-\sigma_8$ constraints from KiDS and DES can be fully reconciled under consistent redshift calibration, thereby suggesting that the different constraints from the two surveys can be traced back to differing methods of redshift calibration.

In tomographic cosmic shear, the signal is primarily sensitive to the average distance of sources within each bin. Therefore, for this purpose, the redshift distribution of an arbitrary galaxy sample can be characterised simply by its mean $\avz$, defined as:
\begin{equation}\label{eq:meanz}
\avz \; = \; \frac{\int_0^\infty \; z \; N(z) \; {\rm d} z}{\int_0^\infty \; N(z) \; {\rm d} z},
\end{equation}
where $N(z)$ is the true redshift distribution of the sample. 
Furthermore, in cosmic shear tomography it is common to build the required tomographic bins using \photoz{} \citep[see][for a review]{salvato19}, which can be measured for large samples of galaxies with observations in only a few photometric bandpasses. However these \photoz{} are imperfect (due to, for example, photometric noise), resulting in tomographic bins whose true $N(z)$ extend beyond the bin limits. These `tails' in the redshift distribution are important, as they can significantly influence the distribution mean and bring sensitive information \citep{ma06}. For a \Euclid{}-like cosmic shear survey, \citet{laureijs11} predict that the mean redshift $\avz$ of each tomographic bin must be known with an accuracy better than $\sigma_{\avz} = 0.002 \,(1+z)$ in order to meet the precision on $w_0$ ($\sigma_{w_0} = 0.015$) and $w_a$ ($\sigma_{w_a} = 0.15$).

Given the importance of measuring the mean redshift for cosmic-shear surveys, numerous approaches have been devised in the last decade. A first family of methods, usually referred to as `direct calibration', involves weighting a sample of galaxies with known redshifts such that they match the colour-magnitude properties of the target galaxy sample; thereby leveraging the relationship between galaxy colours, magnitudes, and redshifts to reconstruct the redshift distribution of the target sample \citep[e.g.][]{lima08, cunha09, abdalla08}. A second approach is to utilise redshift probability distribution functions (\zPDF{}s), obtained per target galaxy and subsequently stacked them to reconstruct the target population $N(z)$. The galaxy \zPDF{} is typically estimated by either model fitting or via machine learning. A third family of methods uses galaxy spatial information, specifically galaxy angular clustering, cross-correlating target galaxies with a large \specz{} sample to retrieve the redshift distribution \citep[e.g.][]{newman08, menard13}. New methods are continuously developed, for instance by modelling galaxy populations and using forward modelling to match the data \citep[][]{Kacprzak20}.

In this paper we evaluate our capacity to measure the mean redshift in each tomographic bin at the precision level required for \Euclid{}, based on realistic simulations.

We base our study on a mock catalogue generated from the Horizon-AGN hydrodynamical simulation as described in \citet{dubois14} and \citet{laigle19}. The advantage of this simulation is that the produced spectra encompass all the complexity of galaxy evolution, including rapidly varying star-formation histories, metallicity enrichment, mergers, and feedback from both supernovae and active galactic nuclei (AGN). By simulating galaxies with the imaging sensitivity expected for \Euclid{}, we retrieve the \photoz{} with a standard template-fitting code, as done in existing surveys. Therefore, we produce \photoz{} with realistic biases, precision and failure rate, as shown in \citet{laigle19}. The simulated galaxy \zPDF{} appear as complex as the ones observed in real data.

We further simulate realistic spectroscopic training samples, with selection functions similar with those that are currently being acquired in preparation of \Euclid{} and other dark energy experiments \citep{masters17}. We introduce possible incompleteness and failures as occurring in actual spectroscopic surveys.

We investigate two of the methods envisioned for the \Euclid{} mission: the direct calibration and \zPDF{} combination. We also propose a new method to debias the \zPDF{} based on  \citet{bordoloi10}. We quantify their performance to estimate the mean redshift of tomographic bins, and isolate relevant factors which could impact our ability to fulfill the \Euclid{} requirement. We also provide recommendations on the imaging depth and training sample necessary to achieve the required accuracy on $\avz{}$.

Finally, we demonstrate the general utility of each of the methods presented here, not just to future surveys such as \Euclid{} but also to current large imaging surveys. As an illustration, we apply those methods to COSMOS and the fourth data release of KiDS \citep[][]{Kuijken+19} surveys.

The paper is organised as follows. In Sect.~\ref{sec:simulation} we describe the \Euclid{}-like mock catalogues generated from the Horizon-AGN hydrodynamical simulation. In Sect.~\ref{sec:direct} we test the precision reached on $\avz$ when applying the direct calibration method. In Sect.~\ref{sec:PDF} we measure $\avz$ in each tomographic bin using the \zPDF{} debiasing technique. We discuss the advantages and limitations of both methods in Sect.~\ref{sec:discussion}. We apply these methods to the KiDS and COSMOS data set in Sect.~\ref{sec:data}.
Finally, we summarise our findings and provide closing remarks in Sect.~\ref{sec:summary}.


\section{A \Euclid{} mock catalogue}
\label{sec:simulation}

In this section we present the \Euclid{} mock catalogue used in this analysis, which is constructed from the Horizon-AGN hydrodynamical simulated lightcone and includes photometry and photometric redshift information. A full description of this mock catalogue can be found in \citet{laigle19}. Here we summarise its main features 
and discuss the construction of several simulated spectroscopic samples, which reproduce a number of expected spectroscopic selection effects. 

\subsection{Horizon-AGN simulation}
\label{subsec:horizon}

Horizon-AGN is a cosmological hydrodynamical simulation ran in a simulation box of 100 $h^{-1}$Mpc per-side, and with a dark matter mass resolution of $8\times 10^7 \, \Msolar$ \citep{dubois14}.  A flat $\Lambda$CDM cosmology with $H_{0}=70.4$\,km\,s$^{-1}$\,Mpc$^{-1}$, $\Omega_{\rm m}=0.272$,  $\Omega_{\Lambda}=0.728$,  and $n_{\rm s}=0.967$ \citep[compatible with WMAP-7,][]{komatsu11} is assumed. Gas evolution is followed on an adaptive mesh, whereby an initial coarse 1024$^{3}$ grid is refined down to $1$~physical kpc. The refinement procedure leads to a typical number of $6.5\times 10^9$ gas resolution elements (called leaf cells) in the simulation at $z=1$. Following \citet{haardt&madau96}, heating of the gas by a uniform ultra-violet background radiation field takes place after $z=10$. Gas in the simulation is able to cool down to temperatures of $10^{4}\,{\rm K}$ through H and He collision, and with a contribution from metals as tabulated in \citet{sutherland&dopita93}. Gas is converted into stellar particles in regions where the gas particle number density surpasses $n_0=0.1\, \rm H\, cm^{-3}$, following a Schmidt law, as explained in \cite{dubois14}. Feedback from stellar winds and supernovae (both types Ia and II) are included in the simulation, and include mass, energy, and metal releases. 
Black holes (BHs) in the simulation can grow by gas accretion, at a Bondi accretion rate that is capped at the Eddington limit, and are able to coalesce when they form a sufficiently tight binary. They release energy in either the quasar or radio (i.e. heating or jet) mode, when the accretion rate is respectively above or below one per cent of the Eddington ratio. The efficiency of these energy release modes are tuned to match the observed BH-galaxy scaling relation at $z=0$ \citep[see][for more details]{dubois12}.

The simulation lightcone was extracted as described in \cite{pichon10}. Particles and gas leaf cells were extracted at each time step depending on their proper distance to the observer at the origin. In total, the lightcone contains roughly $22\,000$ portions of concentric shells, which are taken from about 19 replications of the Horizon-AGN box up to $z=4$. We restrict ourselves to the central 1\,deg$^{2}$ of the lightcone. \citet{laigle19} extracted a galaxy catalogue from the stellar particle distribution using the \textsc{AdaptaHOP} halo finder \citep{aubert04}, where galaxy identification is based exclusively on the local stellar particle density. Only galaxies with stellar masses $M_\star \, > \, 10^9 \Msol$ (which corresponds to around 500 stellar particles) are kept in the final catalogue, resulting in more than $7\times 10^{5}$ galaxies in the redshift range $0<z<4$, with a spatial resolution of 1 kpc.

A full description of the per-galaxy spectral energy distribution (SED)  computation within Horizon-AGN is presented in  \cite{laigle19}\footnote{Horizon-AGN photometric catalogues and SEDs can be downloaded from \url{https://www.horizon-simulation.org/data.html}}, in the following we only summarise the key details of the SED construction process. Each stellar particle in the simulation is assumed to behave as a single stellar population, and its contribution to the galaxy spectrum is generated using the stellar population synthesis models of  \citet{bc03}, assuming a \citet{chabrier03} initial mass function. As each galaxy is composed of a large number of stellar particles, the galaxy SEDs therefore naturally capture the complexities of unique star-formation and chemical enrichment histories. Additionally, dust attenuation is also modelled for each star particle individually, using the mass distribution of the gas-phase metals as a proxy for the dust distribution, and adopting a constant dust-to-metal mass ratio. Dust attenuation (neglecting scattering) is therefore inherently geometry-dependent in the simulation. Finally, absorption of SED photons by the intergalactic medium (i.e. H{\sc i} absorption in the Lyman-series) is modelled along the line of sight to each galaxy, using our knowledge of the gas density distribution in the lightcone. This therefore introduces variation in the observed intergalactic absorption across individual lines of sight. Flux contamination by nebular emission lines is not included in the simulated SEDs. While emission lines could add some complexity in galaxy's photometry, their contribution could be modelled in template-fitting code. Moreover, their impact is mostly crucial at high redshift \citep[][]{schaerer09_em_lines} and when using medium bands \citep[e.g.][]{ilbert09}.

\citet[][]{kaviraj17} compare the global properties of the simulated galaxies with statistical measurements available in the literature (as the luminosity functions, the star-forming main sequence, or the mass functions). They find an overall fairly good agreement with observations. Still, the simulation over-predicts the density of low-mass galaxies, and the median specific star formation rate falls slightly below the literature results, a common trend in current simulations.

\begin{figure}
    \centering
    \includegraphics[width=0.96\columnwidth]{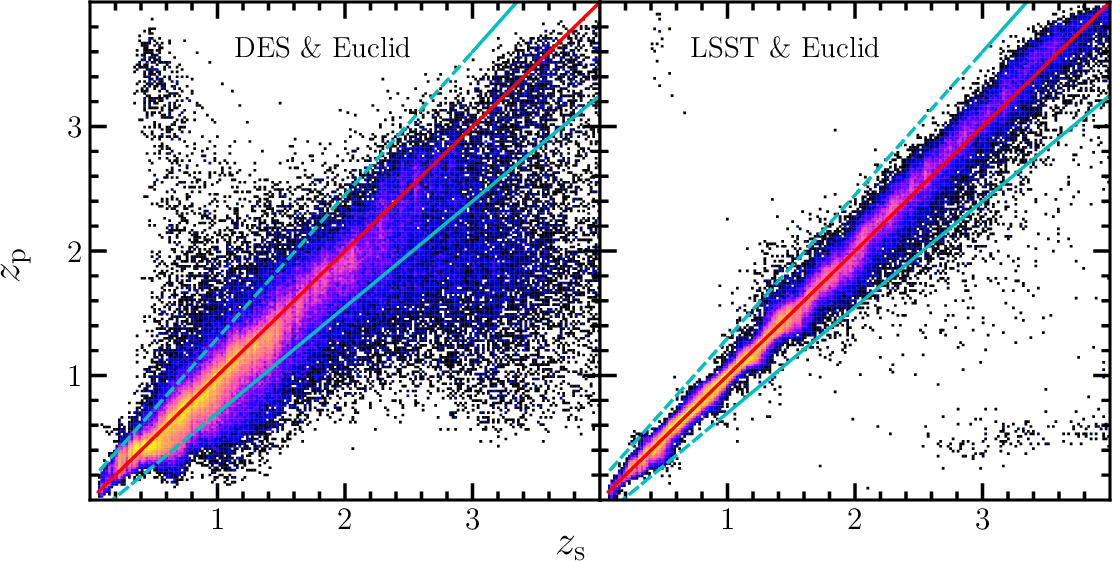}
    \caption{Comparison between the photometric redshifts ($z_{\rm p}$) and spectroscopic redshifts ($z_{\rm s}$) for the Horizon-AGN simulated galaxy sample. Each panel shows a two-dimensional histogram with logarithmic colour scaling, and is annotated with both the 1:1 equivalence line (red) and $|z_{\rm p}-z_{\rm s}|=0.15\,\left(1+z_{\rm s}\right)$ outlier thresholds (blue), for reference. Photometric redshifts are computed using both DES/\Euclid{} (left) and LSST/\Euclid{} (right) simulated photometry, assuming a \Euclid{}-based magnitude limited sample with \RIZ$<24.5$.}
    \label{fig:zmlzs}
\end{figure}

\begin{figure*}
\begin{tabular}{l l l l}
\includegraphics[width=4.2cm]{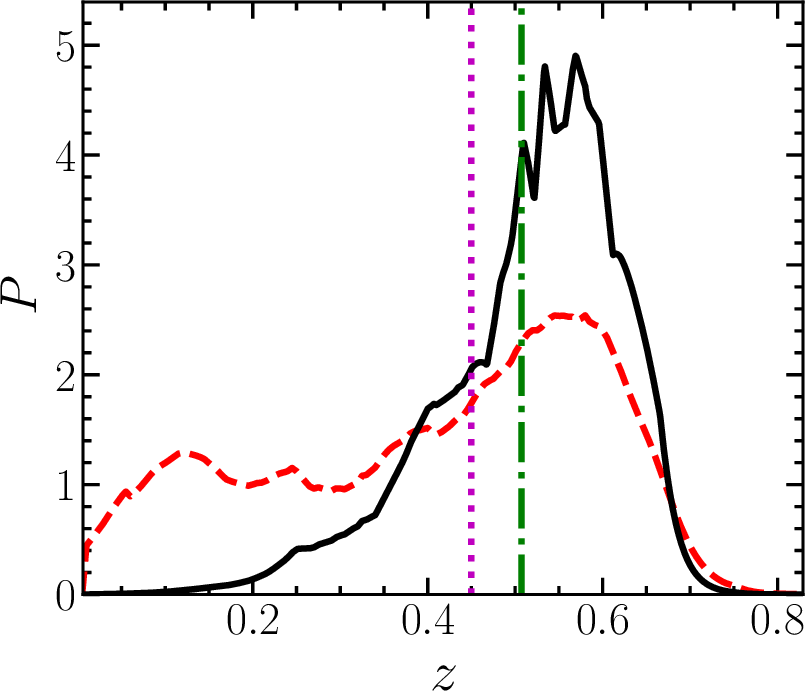}  &
\includegraphics[width=4.2cm]{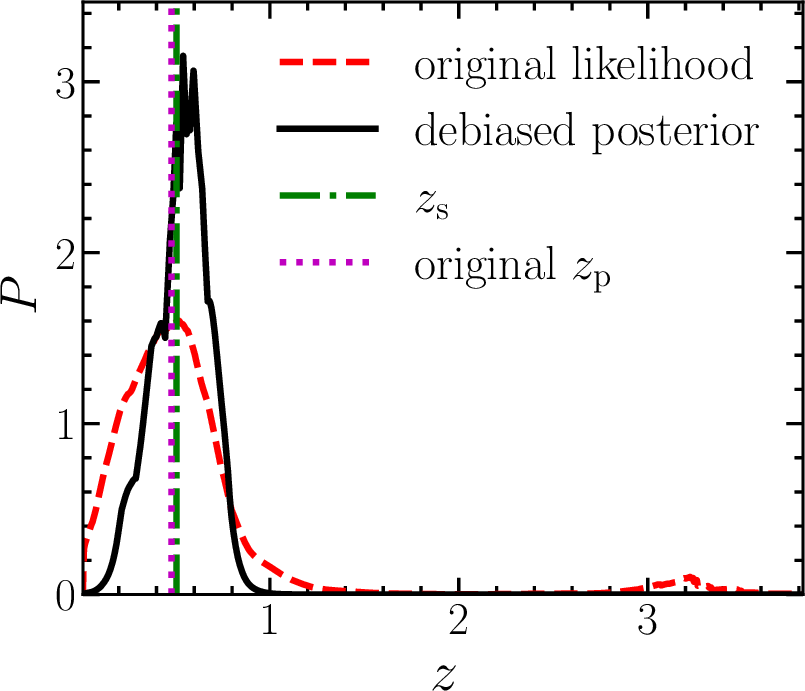}  &
\includegraphics[width=4.2cm]{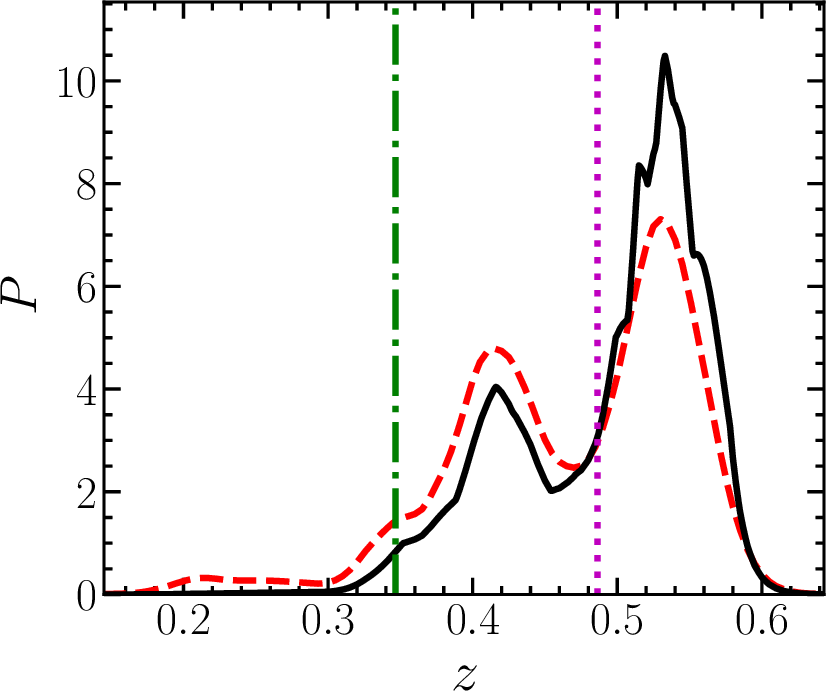}  &
\includegraphics[width=4.2cm]{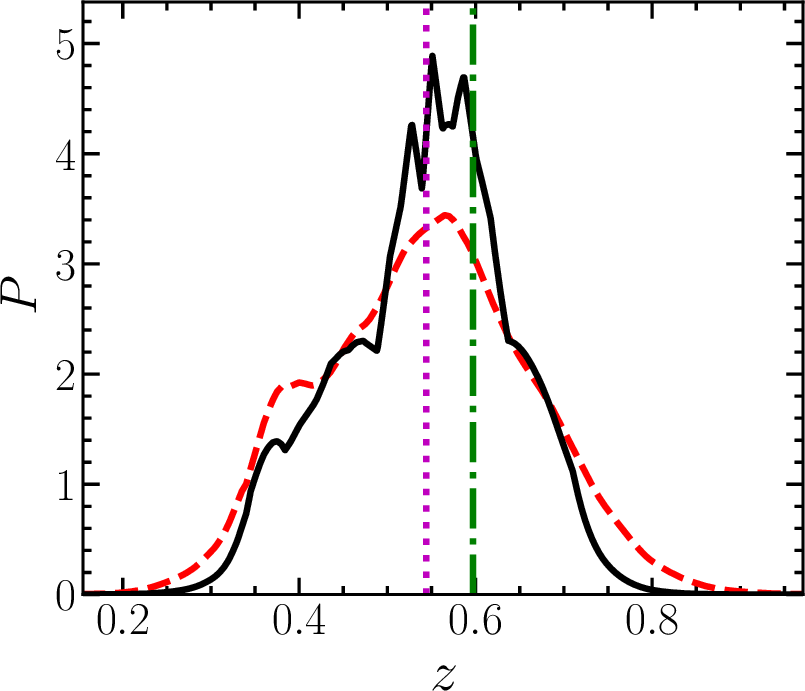} \\   
\includegraphics[width=4.2cm]{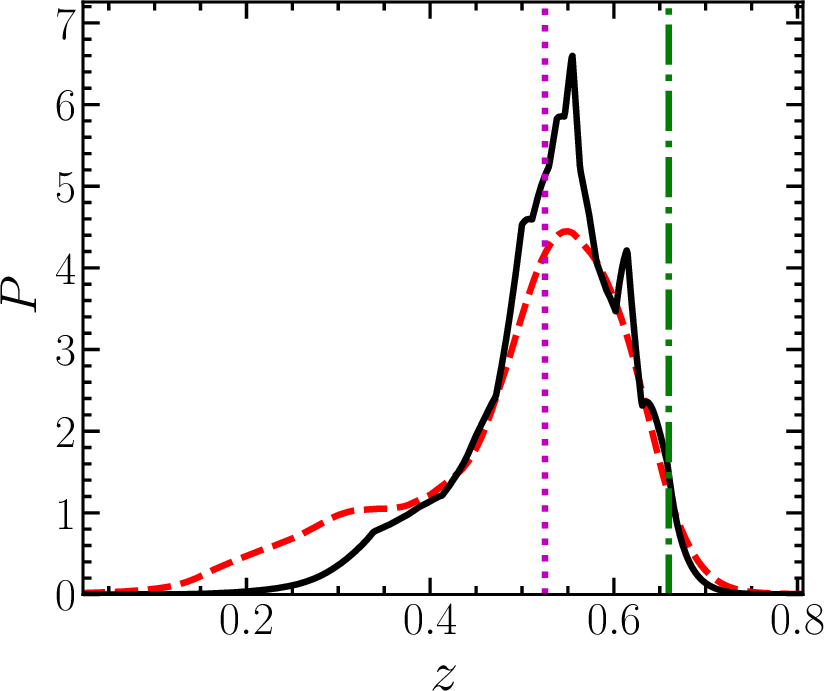}  &
\includegraphics[width=4.2cm]{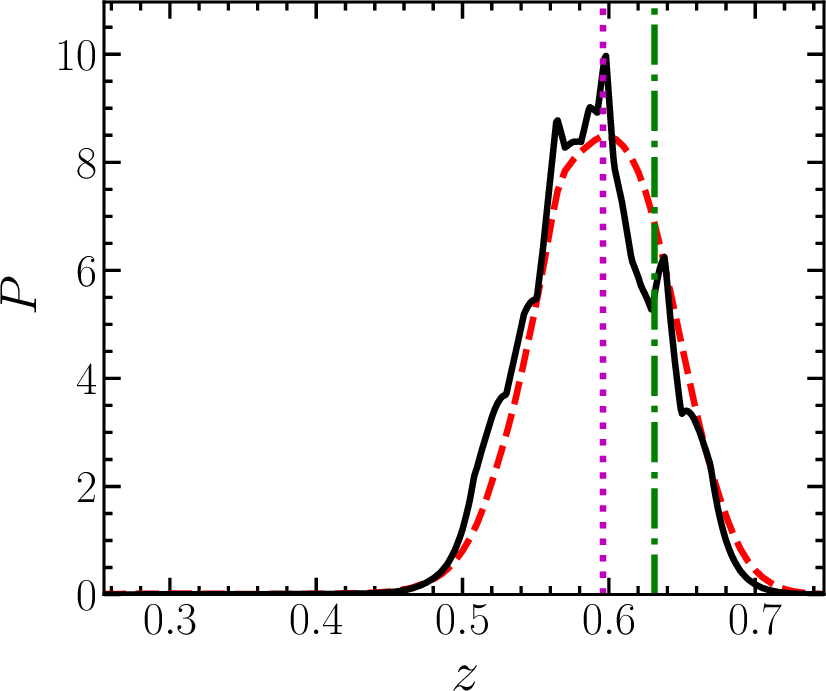} &   
\includegraphics[width=4.2cm]{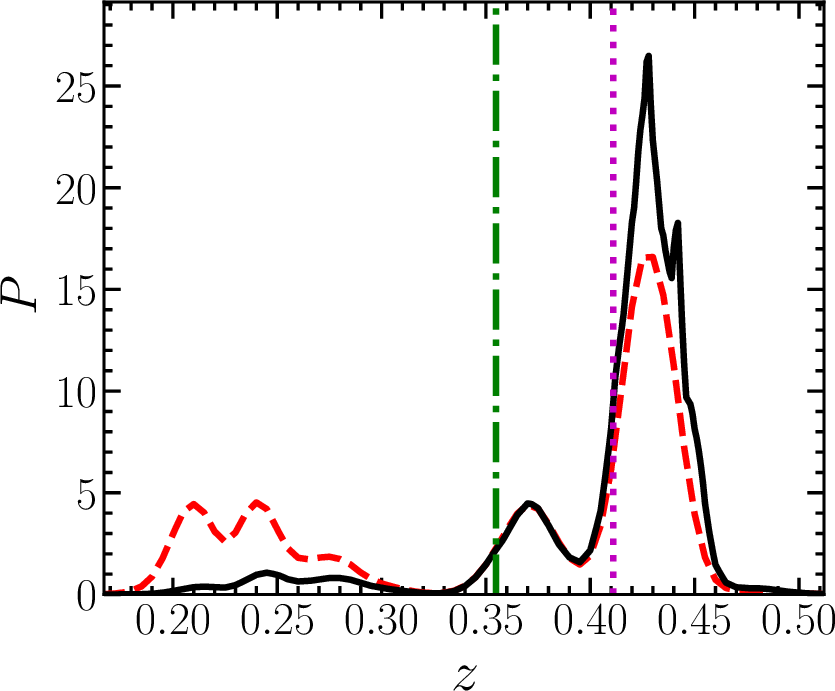}  &
\includegraphics[width=4.2cm]{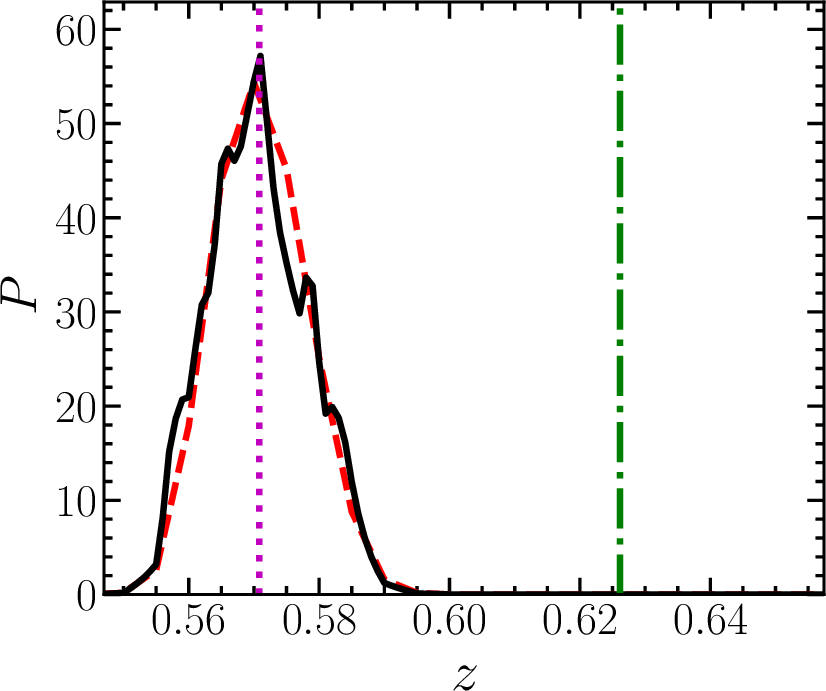} 
\end{tabular}
\caption{Few examples of galaxy likelihood ${\mathcal L }(z)$ (dashed red lines) and debiased posterior distributions (solid black lines). The \specz{} (\photoz{}) are indicated with green (magenta) dotted lines. These galaxies are selected in the tomographic bin $0.4<z_{\rm p}<0.6$ for the DES/\Euclid{} (top panels) and LSST/\Euclid{} (bottom panels) configurations. These likelihoods are not a random selection of sources, but illustrate the variety of likelihoods present in the simulations.}
\label{fig:PDFexamples}
\end{figure*}

\subsection{Simulation of \Euclid{} photometry and photometric redshifts}
\label{subsec:photoz}

As described in \citet{laureijs11}, the \Euclid{} mission will measure the shapes of about $1.5$ billion galaxies over $15\,000$ deg$^2$. The visible (VIS) instrument will obtain images taken in one very broad filter (\RIZ{}), spanning 3500 \AA{}. This filter allows extremely efficient light collection, and will enable VIS to measure the shapes of galaxies as faint as $24.5$ mag with high precision. The near infrared spectrometer and photometer (NISP) instrument will produce images in three near-infrared (NIR) filters. In addition to these data, \Euclid{} satellite observations are expected to be complemented by large samples of ground-based imaging, primarily in the optical, to assist the measurement of \photoz{}.

\Euclid{} imaging has an expected sensitivity, over $15\,000$ deg$^2$, of 24.5 mag  (at $10\sigma$) in the $VIS$ band, and 24 mag (at $5\sigma$) in each of the $Y$, $J$, and $H$ bands \citep{laureijs11}. We associate the \Euclid{} imaging with two possible ground-based visible imaging datasets, which correspond to two limiting cases for \photoz{} estimation performance.
\begin{itemize}
    \item \textbf{DES/\Euclid{}}. As a demonstration of \photoz{} performance when combining \Euclid{} with a considerably shallower photometric dataset, we combine our \Euclid{} photometry with that from DES \citep{abbott18}. DES imaging is taken in the $g$, $r$, $i$, and $z$ filters, at $10\sigma$ sensitivities of $24.33$, $24.08$, $23.44$, and $22.69$ respectively.
    \item \textbf{LSST/\Euclid{}}. As a demonstration of \photoz{} performance when combining \Euclid{} with a considerably deeper photometric dataset, we combine our \Euclid{} photometry with that from the Vera C. Rubin Observatory LSST \citep[][]{LSSTsciencebook}. LSST imaging will be taken in the $u$, $g$, $r$, $i$, $z$, and $y$ filters, at $5\sigma$ (point source, full depth) sensitivities of $26.3$, $27.5$, $27.7$, $27.0$, $26.2$, and $24.9$, respectively.
\end{itemize}
DES imaging is completed and meets these expected sensitivities. Conversely LSST will not reach those quoted full depth sensitivities before its tenth year of operation (starting in 2021), and even then it is possible that the northern extension of LSST might not reach the same depth. Still, LSST will be already extremely deep after two years of operation, being only 0.9 magnitude shallower than the final expected sensitivity \citep[][]{graham20_LSST}. Therefore, these two cases (and their assumed sensitivities) should comfortably encompass the possible \photoz{} performance of any future combined optical and \Euclid{} photometric data set.

In order to generate the mock photometry in each of the \Euclid{}, DES, and LSST surveys, each galaxy SED is first `observed' through the relevant filter response curves. In each photometric band, we generate Gaussian distributions of the expected signal-to-noise ratios (SNs) as a function of magnitude, given both the depth of the survey and typical SN-magnitude relation (in the same wavelength range) \citep[see appendix A in][]{laigle19}. We then use these distributions, per filter, to assign each galaxy a SN (given its magnitude). The SN of each galaxy determines its `true' flux uncertainty, which is then used to perturb the photometry (assuming Gaussian random noise) and produce the final flux estimate per source. This process is then repeated for all desired filters.

The galaxy \photoz{} are derived in the same manner as with real-world photometry. We use the method detailed in \citet{ilbert13}, based on the template-fitting code \lephare{} \citep{arnouts02,ilbert06}. We adopt a set of 33 templates from \citet{polletta07} complemented with templates from \citet{bc03}. Two dust attenuation curves are considered \citep{prevot84,calzetti2000}, allowing for a possible bump at 2175\AA. Neither emission lines nor adaptation of the zero-points are considered, since they are not included in the simulated galaxy catalogue. The full redshift likelihood, ${\mathcal L }(z)$, is stored for each galaxy, and the \photoz{} point-estimate, $z_{\rm p}$, is defined as the median of ${\mathcal L }(z)$\footnote{The median of ${\mathcal L }(z)$ could differ from the peak of ${\mathcal L }(z)$, or from the redshift corresponding to the minimum $\chi^2$, especially for ill-defined likelihoods.}. The distributions of  (derived) photometric redshift versus (intrinsic) spectroscopic redshift for mock galaxies (in both our DES/\Euclid{} and LSST/\Euclid{} configurations) are shown in Fig.~\ref{fig:zmlzs}. Several examples of redshift likelihoods are shown in Fig.~\ref{fig:PDFexamples}. We can see realistic cases with multiple modes in the distribution, as well as asymmetric distributions around the main mode. The \photoz{} used to select galaxies within the tomographic bins are indicated by the magenta lines and that they can differ significantly from the \specz{} (green lines). 

We wish to remove galaxies with a broad likelihood distribution (i.e. galaxies with truly uncertain \photoz{}) from our sample. In practice, we approximate the breadth of the likelihood distribution using the \photoz{} uncertainties produced by the template-fitting procedure to clean the sample. \lephare{} produces a redshift confidence interval $[z_{\rm p}^{\rm min},z_{\rm p}^{\rm max}]$, per source, which encompasses 68\% of the redshift probability around $z_{\rm p}$. We remove galaxies with $\max( \, z_{\rm p}-z_{\rm p}^{\rm min} \, , \, z_{\rm p}^{\rm max}-z_{\rm p}\,) \,> \, 0.3$, which we denote $\sigma_{z_{\rm p}}>0.3$ in the following for simplicity. We investigate the impact of this choice on the number of galaxies available for cosmic shear analyses, and also quantify the impact of relaxing this limit, in Sect.~\ref{subsec:discussionErr}.

Finally, we generate 18 photometric noise realisations of the mock galaxy catalogue. While the intrinsic physical properties of the simulated galaxies remain the same under each of these realisations,
the differing photometric noise allows us to quantify the role of photometric noise alone on our estimated of $\avz{}$. We only adopt 18 realisations due to computational limitations, however, our results are stable to the addition of more realisations.

\subsection{Definition of the target photometric sample and the spectroscopic training samples}
\label{subsec:specDescription}

All redshift-calibration approaches discussed in this paper utilise a \specz{} training sample to estimate the mean redshift of a target photometric sample. In practice, such a spectroscopic training sample is rarely a representative subset of the target photometric sample, but is often composed of bluer and brighter galaxies. Therefore, to properly assess the performance of our tested approaches, we must ensure that the simulated training sample is distinct from the photometric sample. To do this, we separate the Horizon-AGN catalogue into two equal sized subsets:  we define the first half of the photometric catalogue as our as target sample, and draw variously defined spectroscopic training samples from the second half of the catalogue. We test each of our calibration approaches with three spectroscopic training samples, designed to mimic different spectroscopic selection functions:
\begin{itemize}
    \item a uniform training sample;
    \item a SOM-based training sample; 
    \item and a COSMOS-like training sample.
\end{itemize}    

The uniform training sample is the simplest, most idealised training sample possible. We sample 1000 galaxies with \RIZ{}$<24.5$ mag (i.e. the same magnitude limit as in the target sample) in each tomographic bin, independently of all other properties. While this sample is ideal in terms of representation, the sample size is set to mimic a realistic training sample that could be obtained from dedicated ground-based spectroscopic follow-up of a \Euclid{}-like target sample.

Our second training sample follows the current \Euclid{} baseline to build a training sample. \citet{masters17} endeavour to construct a spectroscopic survey, the Complete Calibration of the Colour-Redshift Relation survey (C3R2), which
completely samples the colour/magnitude space of cosmic shear target samples. This sample is currently assembled by combining data from ESO and Keck facilities \citep[][]{masters19,guglielmo20_c3r2}. The target selection is based on an unsupervised machine-learning technique, the self-organising map \citep[SOM,][]{Kohonen82}, which they use to define a spectroscopic target sample that is representative in terms of galaxy colours of the \Euclid{} cosmic shear sample. The SOM allows a projection of a multi-dimensional distribution into a lower two-dimensional map. The utility of the SOM lies in its preservation of higher-dimensional topology: neighbouring  objects in the multi-dimensional space fall within similar regions of the resulting map. This allows the SOM to be utilised as a multi-dimensional clustering tool, whereby discrete map cells associate sources within discrete voxels in the higher dimensional space. We utilise the method of \citet{Davidzon19} to construct a SOM, which involves projecting  observed (i.e. noisy) colours of the mock catalogue into a map of 6400 cells (with dimension $80 \times 80$). We construct our SOM using the LSST/\Euclid{} simulated colours, assuming implicitly that the \specz{} training sample is defined using deep calibration fields. If the flux uncertainty is too large ($\Delta m^x_i>0.5$, for object $i$ in filter $x$) the observed magnitude is replaced by that predicted from the best-fit SED template, which is estimated while preparing the SOM input catalogue. This procedure allows us to retain sources that have non-detections in some photometric bands. We then construct our SOM-based training sample by randomly selecting $N_{\rm train}$ galaxies from each cell in the SOM. The C3R2 expects to have $\geqslant 1$ spectroscopic galaxies per SOM cell available for calibration by the time that the \Euclid{} mission is active. For our default SOM coverage, we invoke a slightly more idealised situation of two galaxies per cell and we impose that these two galaxies belong to the considered tomographic bin. This procedure ensures that all cells are represented in the spectroscopy. In reality, a fraction of cells will likely not contain spectroscopy. However, when treated correctly, such misrepresented cells act only to decrease the target sample number density, and do not bias the resulting redshift distribution mean estimates \citep{wright20}. We therefore expect that this idealised treatment will not produce results that are overly-optimistic. 

Finally, the COSMOS-like training sample mimics a typical heterogeneous spectroscopic sample, currently available in the COSMOS field. We first simulate the zCOSMOS-like spectroscopic sample \citep{lilly07}, which consists of two distinct components: a bright and a faint survey. The zCOSMOS-Bright sample is selected such that it contains only galaxies at $z<1.2$, while the zCOSMOS-Faint sample contains only galaxies at $z>1.7$ (with a strong bias towards selecting  star-forming galaxies). To mimic these selections, we construct a mock sample whereby half of the sources are brighter than $i=22.5$ (the bright sample) and half of the galaxies reside at  $1.7<z<2.4$ with $g<25$ (the faint sample). We then add to this compilation a sample of $2000$ galaxies that are randomly selected at $i<25$, mimicking the low-z VUDS sample \citep{lefevre15}, and a sample of $1000$ galaxies randomly selected at $0.8<z<1.6$ with $i<24$, mimicking the sample of \citet{comparat15}. By construction, this final spectroscopic redshift compilation exhibits low representation of the photometric target sample in the redshift range $1.3<z<1.7$.

Overall, our three training samples exhibit (by design) differing redshift distributions and galaxy number densities. We investigate the sensitivity of the estimated $\avz$ on the size of the training sample in Sect.~\ref{subsec:size}.

\begin{figure}
    \centering
    \includegraphics[width=0.96\columnwidth]{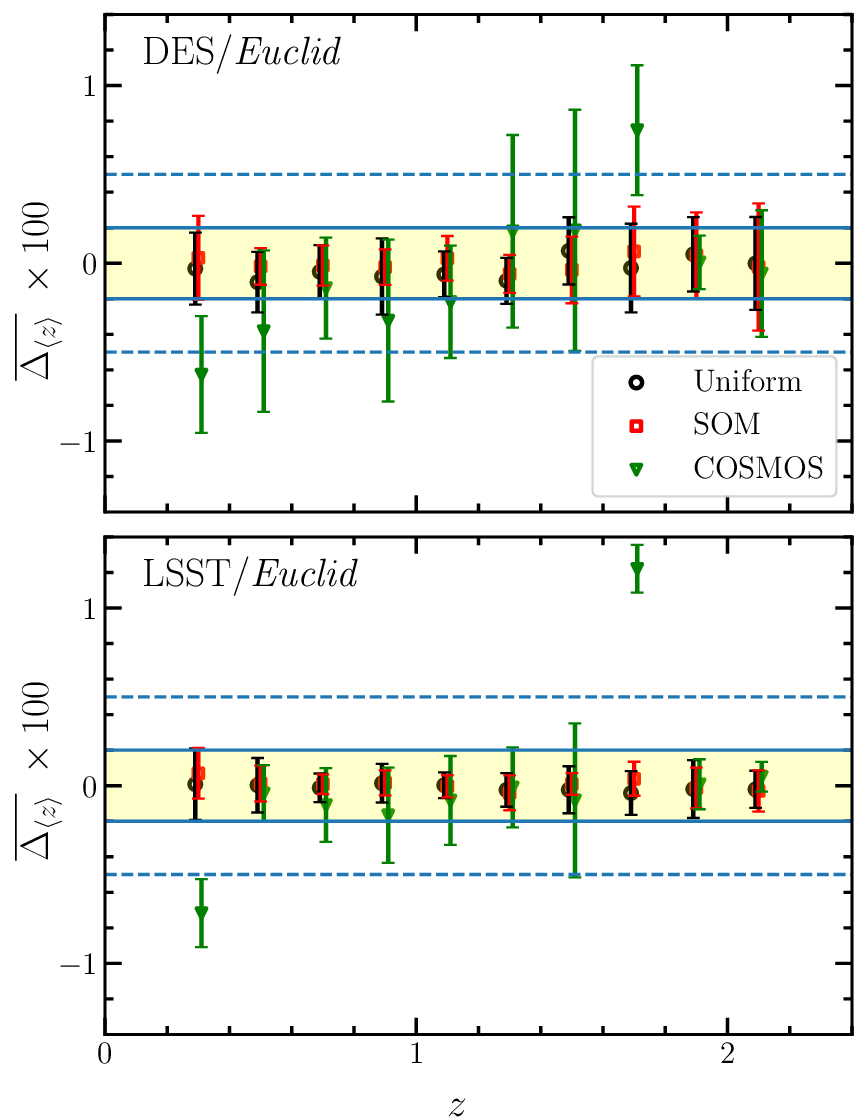}
    \caption{Bias on the mean redshift (see Eq.~\ref{eq:deltaz}) averaged over the 18 photometric noise realisations. The mean redshifts are measured using the direct calibration approach. The tomographic bins are defined using the DES/\Euclid{} and LSST/\Euclid{} \photoz{} in the top and bottom panels, respectively. The yellow region represents the \Euclid{} requirement at $0.002\,(1+z)$ for the mean redshift accuracy, and the blue dashed lines correspond to a bias of $0.005\,(1+z)$. The symbols represent the results obtained with different training samples: (a) selecting uniformly 1000 galaxies per tomographic bin (black circles); (b) selecting two galaxies/cell in the SOM (red squares); and (c) selecting a sample that mimics real spectroscopic survey compilations in the COSMOS field (green triangles). }
    \label{fig:direct}
\end{figure} 

\section{Direct calibration}
\label{sec:direct}

Direct calibration is a fairly straightforward method that can be used to estimate the mean redshift of a photometric galaxy sample, and is currently the baseline method planned for \Euclid{} cosmic shear analyses. In this section we describe our implementation of the direct calibration method, apply this method to our various spectroscopic training samples, and report the resulting accuracy of our redshift distribution mean estimates. 

\subsection{Implementation for the different training samples}
Given our different classes of training samples, we are able to implement slightly different methods of direct calibration. We detail here how the implementation of direct calibration differs for each of our three spectroscopic training samples.

{\bf The uniform sample.} In the case where the training sample is known to uniformly sparse-sample the target galaxy distribution, an estimate of $\avz$ can be approximated by simply computing the mean redshift of the training sample. 

{\bf The SOM sample.} By construction, the SOM training sample uniformly covers the full n-dimensional colour space of the target sample. 
The method relies on the assumption that galaxies within a cell share the same redshift \citep[][]{masters15} which can be labelled with the training sample. Therefore, we can estimate the mean redshift of the target distribution $\avz$ by simply calculating the weighted mean of each cell's average redshift, where the weight is the number of target galaxies per cell: 
\begin{equation}\label{eq:zmeanSOM}
    \avz = \frac{1}{N_{\rm t}} \; \sum_{i=1}^{N_{\rm cells}} \; \left<z_{\rm train}^i\right> \; N_i,
\end{equation}
where the sum runs over the $i\in [1,N_{\rm cells}]$ cells in the SOM, $\left<z_{\rm train}^i\right>$ is the mean redshift of the training spectroscopic sources in cell $i$,  $N_i$ is the number of target galaxies (per tomographic bin) in cell $i$, and $N_{\rm t}$ is the total number of target galaxies in the tomographic bin. A shear weight associated to each galaxy can be introduced in this equation \citep[e.g.][]{wright20}. As described in Sect.~\ref{subsec:specDescription}, our SOM is consistently constructed by training on LSST/\Euclid{} photometry, even when studying the shallower DES/\Euclid{} configuration. We adopt this strategy since the training spectroscopic samples in \Euclid{} will be acquired in calibration fields \citep[e.g.][]{masters19} with deep dedicated imaging. This assumption implies that the target distribution $\avz{}$ is estimated exclusively in these calibration fields, which are covered with photometry from both our shallow and deep setups, and therefore increases the influence of sample variance on the calibration.

{\bf The COSMOS-like sample.} Applying direct calibration to a heterogeneous training sample is less straightforward than in the above cases, as the training sample is not representative of the target sample in any respect. Weighting of the spectroscopic sample, therefore, must correct for the mix of spectroscopic selection effects present in the training sample, as a function of magnitude (from the various magnitude limits of the individual spectroscopic surveys), colour (from their various preselections in colour and spectral type), and redshift (from dedicated redshift preselection, such as that in zCOSMOS-Faint). Such a weighting scheme can be established efficiently with machine-learning techniques such as the SOM. To perform this weighting, we train a new SOM using all the information that have the potential to correct for the selection effects present in our heterogeneous training sample: apparent magnitudes, colours, and template-based \photoz{}. We create this SOM using only the galaxies from the COSMOS-like sample that belong to the considered tomographic bin, and reduce the size of the map to 400 cells ($20\times20$, because the tomographic bin itself spans a smaller colour space). Finally, we project the target sample into the SOM and derive weights for each training sample galaxy, such that they reproduce the per-cell density of target sample galaxies. This process follows the same weighting procedure as \citet{wright20}, who extend the direct calibration method of \citet{lima08} to include source groupings defined via the SOM. In this method, the estimate of $\avz$ is also inferred using Eq.~(\ref{eq:zmeanSOM}).

\subsection{Results}
\label{subsec:results}

We apply the direct calibration technique to the mock catalogue, split into ten tomographic bins spanning the redshift interval $0.2 < z_{\rm p} < 2.2$. To construct the samples within each tomographic bin, training and target samples are selected based on their best-estimate \photoz{}, $z_{\rm p}$. We quantify the performance of the redshift calibration procedure using the measured bias in $\avz$, defined as:
\begin{equation}\label{eq:deltaz}
    \Delta_{\avz} = \frac{\avz-\avz^{\rm true}}{1+\avz^{\rm true}},
\end{equation}
and evaluated over the target sample. We present the values of $\Delta_{\avz}$ that we obtain with direct calibration in Fig.~\ref{fig:direct}, for each of the ten tomographic bins. The figure shows, per tomographic bin, the population mean (points) and $68\%$ population scatter (error bars) of $\Delta_{\avz}$ over the 18 photometric noise realisations of our simulation. The solid lines and yellow region indicate the $|\Delta_{\avz}|\,\leq\,2\times10^{-3}$ requirement stipulated by the \Euclid{} mission. Given our limited number of photometric noise realisations, estimating the population mean and scatter directly from the 18 samples is not sufficiently robust for our purposes. We thus use maximum likelihood estimation, assuming Gaussianity of the $\Delta_{\avz}$ distribution, to determine the underlying population mean and the scatter. We define these underlying population statistics as \biaszmean{} and \scatterzmean{} for the mean and the scatter, respectively. 

We find that, when using a uniform or SOM training sample, direct calibration is consistently able to recover the target sample mean redshift to $|\biaszmeanB{}|\,<\,2\times10^{-3}$. In the case of the shallow DES/\Euclid{} configuration, however, the scatter \scatterzmean{} exceeds the \Euclid{} accuracy requirement in the highest and lowest tomographic bins. The DES/\Euclid{} configuration is, therefore, technically unable to meet the \Euclid{} precision requirement on $\avz$ in the extreme bins. In the LSST/\Euclid{} configuration, conversely, the precision and accuracy requirements are both consistently satisfied. We hypothesise that this difference stems from the deeper photometry having higher discriminatory power in the tomographic binning itself: the $N(z)$ distribution for each tomographic bin is intrinsically broader for bins defined with shallow photometry, and therefore has the potential to demonstrate greater complexity (such as colour-redshift degeneracies) that reduce the effectiveness of direct calibration. 

The direct calibration with the SOM relies on the assumption that galaxies within a cell share the same redshift \citep[][]{masters15}. Noise and degeneracies in the colour-redshift space introduce a redshift dispersion within the cell which impacts the accuracy of $\avz{}$. Even with the diversity of SED generated with Horizon-AGN, and introducing noise in the photometry, we find that the direct calibration with a SOM sample is sufficient to reach the \Euclid{} requirement.

We find that the COSMOS-like training sample is unable to reach the required accuracy of \Euclid{}. This behaviour is somewhat expected, since the COSMOS-like sample contains selection effects that are not cleanly accessible to the direct calibration weighting procedure. The mean redshift is particularly biased in the bin $1.6<z<1.8$, where there is a dearth of spectra; the \citet{comparat15} sample is limited to $z<1.6$, while the zCOSMOS-Faint sample  resides exclusively at $z>1.7$, thereby leaving the range $1.6<z<1.7$ almost entirely unrepresented. In this circumstance, our SOM-based weighting procedure is insufficient to correct for the heterogeneous selection, leading to bias. This is typical in cases where the training sample is missing certain galaxy populations that are present in the target sample \citep{hartley20}. We note, though, that it may be possible to remove some of this bias via careful quality control during the direct calibration process, such as demonstrated in \citet{wright20}. Whether such quality control would be sufficient to meet the \Euclid{} requirements, however, is uncertain. 

We note that, although we are utilising photometric noise realisations in our estimates of $\avz$, the underlying mock catalogue remains the same. As a result, our estimates of \biaszmean{} and \scatterzmean{} are not impacted by sample variance. In reality, sample variance affects the performance of the direct calibration, particularly when assuming that the training sample is directly representative of the target distribution (as we do with our uniform training sample). For fields smaller than 2 deg$^2$, \citet{bordoloi10} showed that Poisson noise dominates over sample variance (in mean redshift estimation) when the training sample consists of less than 100 galaxies. Above this size, sample variance dominates the calibration uncertainty. This means that, in order to generate an unbiased estimate of $\avz$ using a uniform sample of $1000$ galaxies, a minimum of 10 fields of 2 deg$^2$ would need to be surveyed. 

The SOM approach is less sensitive to sample variance, as over-densities (and under-densities) in the target sample population relative to the training sample are essentially removed in the weighting procedure \citep[provided that the population is present in the training sample,][]{lima08,wright20}. In the cells corresponding to this over-represented target population, the relative importance of training sample redshifts will be similarly up-weighted, thereby removing any bias in the reconstructed  $N(z)$. Therefore, sample variance should have only a weak impact on the global derived $N(z)$ in this method. Nonetheless, samples variance may still be problematic if, for example, under-densities result in entire populations being absent from the training sample. 

Finally, it is worth emphasising that these results are obtained assuming perfect knowledge of training set redshifts. We study the impact of failures in spectroscopic redshift estimation in Sect.~\ref{sec:discussion}.

\begin{figure*}
\begin{tabular}{l l}
\includegraphics[width=8.5cm]{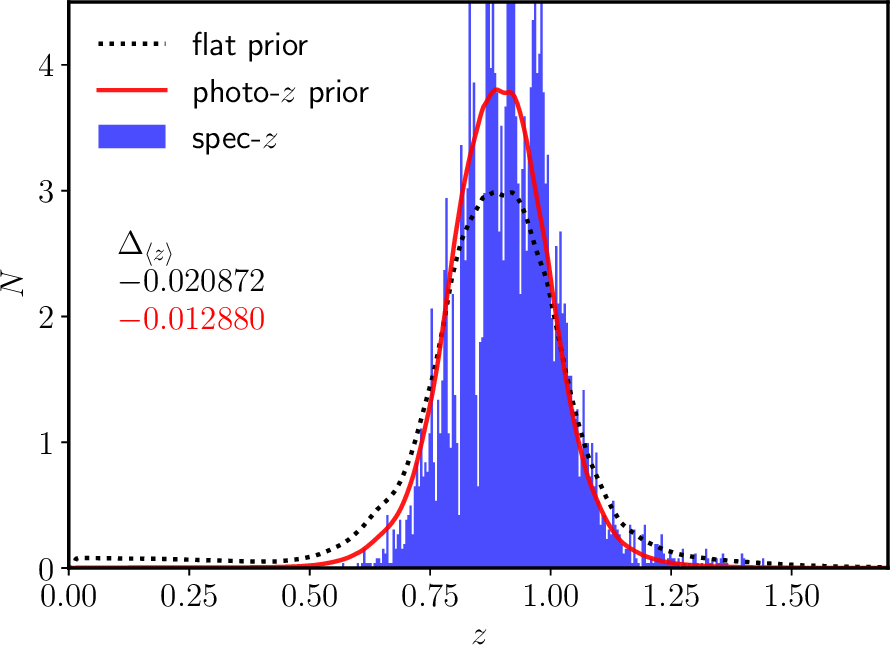}  &
\includegraphics[width=8.5cm]{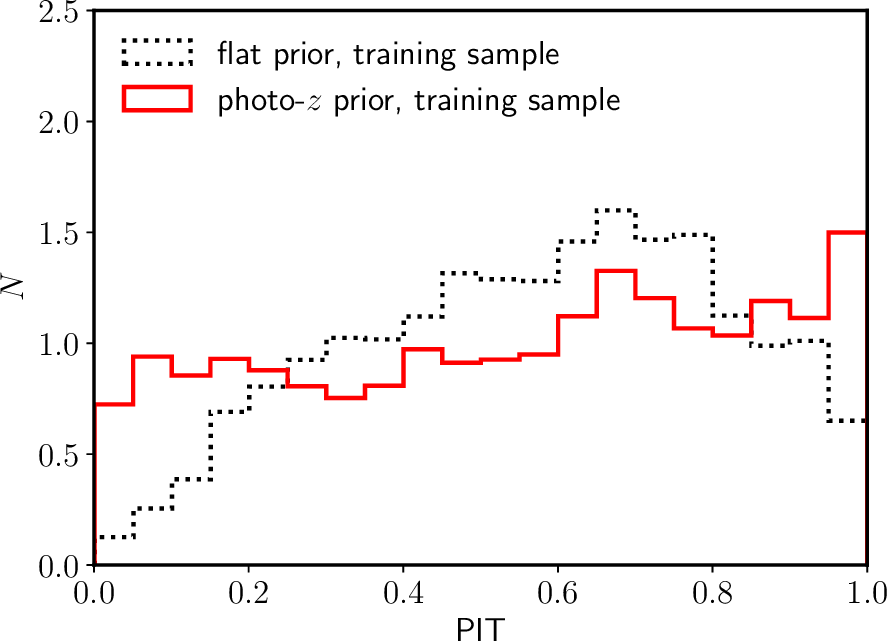} \\   
\includegraphics[width=8.5cm]{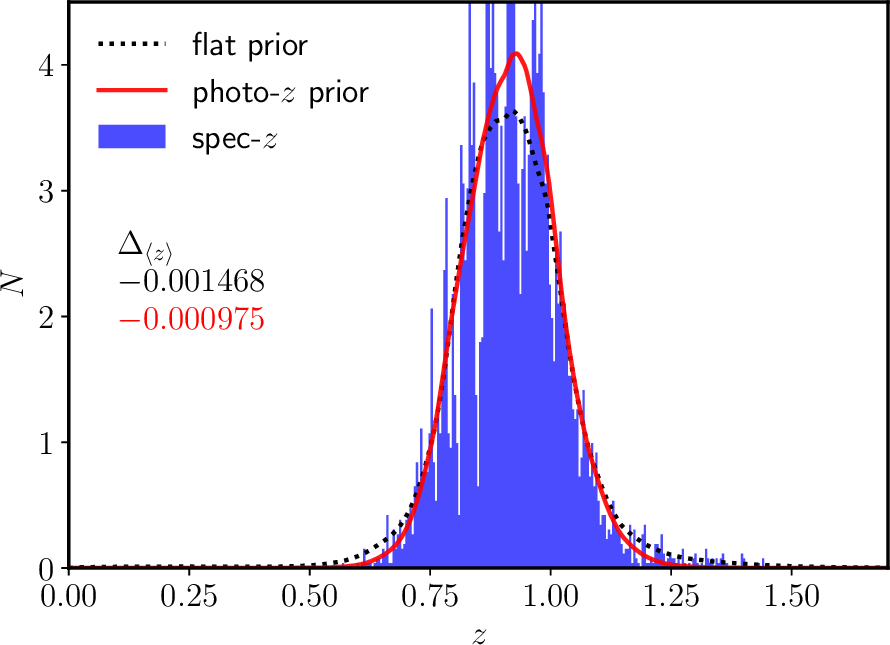}  &
\includegraphics[width=8.5cm]{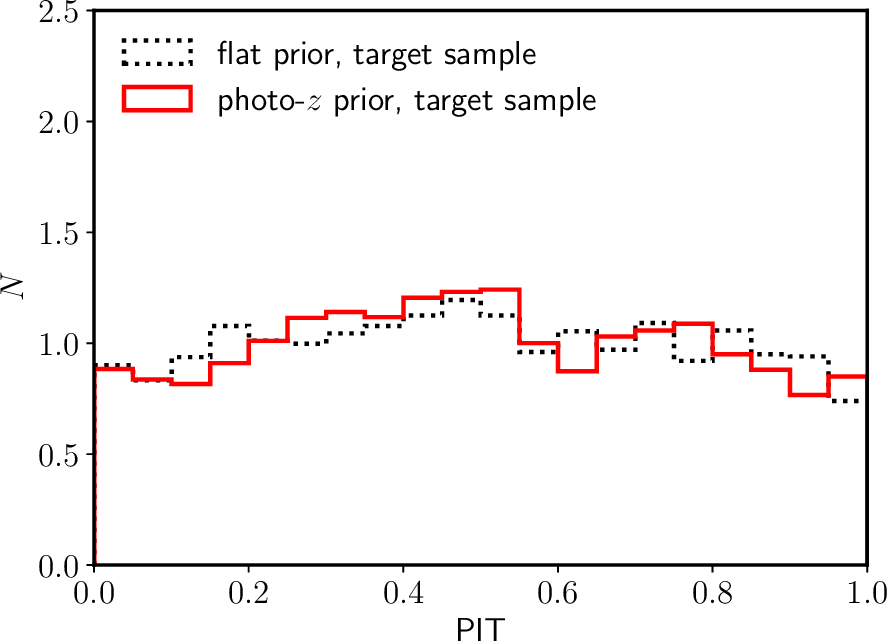} 
\end{tabular}
\caption{Examples of redshift distributions (left) and PIT distributions (right, see text for details) for a tomographic bin selected to $0.8 < z_{\rm p} < 1$ using DES/\Euclid{} \photoz{}. In these examples, we assume a training sample extracted from a SOM, with two galaxies per cell. The top and bottom panels show the results before and after \zPDF{} debiasing, respectively. Redshift distributions and PITs are shown for the true redshift distribution (blue), and redshift distributions estimated using the \zPDF{} method, when incorporating \photoz{} (red) and uniform (black) priors. 
\label{fig:PIT}}
\end{figure*}

\section{Estimator based on redshift probabilities}
\label{sec:PDF}

In this section we present another approach to redshift distribution calibration that uses the information contained in the galaxy redshift probability distribution function, available for each individual galaxy of the target sample. Photometric redshift estimation codes typically provide approximations to this distribution based solely on the available photometry of each source. We study the performance of methods utilising this information in the context of \Euclid{} and test a method to debias the \zPDF{}.

\subsection{Formalism}

Given the relationship between galaxy magnitudes and colours (denoted $\obs$) and redshift $z$, one can utilise the conditional probability $p(z|\obs)$ to estimate the true redshift distribution $N(z)$, using an estimator such as that of  \citet{sheth07,sheth10}:
\begin{equation} \label{eq:nzpdz}
    N(z) = \int N(\obs) \, p(z|\obs) \, {\rm d} \obs = \sum_i^{N_{\rm t}} p_i(z|\obs),
\end{equation}
where $N(\obs)$ is the joint n-dimensional distribution of colours and magnitudes. As made explicit in the above equation, the $N(z)$ estimator reduces simply to the sum of the individual (per-galaxy) conditional redshift probability distributions, $p_i(z|\obs)$. A shear weight associated to each galaxy can be introduced in this equation \citep[e.g.][]{wright20}. It is worth noting that this summation over conditional probabilities is ideologically similar to the summation of SOM-cell redshift distributions presented previously; in both cases, one effectively builds an estimate of the probability $p(z|\obs)$, and uses this to estimate $\avz$. Indeed, it is clear that the SOM-based estimate of $\avz$ presented in  Eq.~(\ref{eq:zmeanSOM}) in fact follows directly from Eq.~(\ref{eq:nzpdz}).

Generally, photometric redshift codes provide in output a normalised likelihood function that gives the probability of the
observed photometry given the true redshift, ${\mathcal L }(\obs|z)$, or sometimes the posterior probability distribution ${\mathcal P }(z|\obs)$ \citep[e.g.][]{benitez00,bolzonella00,arnouts02,cunha09}. These two probability distribution functions are related through the Bayes theorem as,
\begin{equation}
{\mathcal P}(z|\obs) \propto {\mathcal L }(\obs|z) \, {\rm Pr}(z),
\end{equation}
where ${\rm Pr}(z)$ is the prior probability.

Photometric redshift methods that invoke template-fitting, such as the \lephare{} \photoz{} estimation code, generally explore the likelihood of the observed photometry given a range of theoretical templates $T$ and true redshifts ${\mathcal L}(\obs|T,z)$. The full likelihood, ${\mathcal L}(\obs|z)$, is then obtained by marginalising over the template set:
\begin{equation}\label{eq:likedef}
    {\mathcal L}(\obs|z) = \sum_T  {\mathcal L}(\obs|T,z).
\end{equation}
In the full Bayesian framework, however, we are instead interested in the posterior probability, rather than the likelihood. In the formulation of this posterior, we first make explicit the dependence between galaxy  colours $\vec{c}$ and magnitude in one (reference) band $m_0$:  $\obs=\left\{\vec{c},m_0\right\}$. Following \citet[][]{benitez00} we can then define the posterior probability distribution function: 
\begin{equation} 
{\mathcal P}(z|\vec{c},m_0) \propto \sum_T  {\mathcal L}(\vec{c} |T,z)  \, {\rm Pr}(z|T,m_0)  \, {\rm Pr}(T|m_0),
\end{equation} 
where ${\rm Pr}(z|T,m_0)$ is the prior conditional probability of redshift given a particular galaxy template and reference magnitude, and ${\rm Pr}(T|m_0)$ is the prior conditional probability of each template at a given reference magnitude. 
Under the approximation that the redshift distribution does not depend on the template, and that the template distribution is independent of the magnitude (i.e. the luminosity function does not depend on the SED type), one obtains
\begin{eqnarray}
{\mathcal P}(z|\vec{c},m_0) & \propto & \sum_T  {\mathcal L}(\vec{c} |T,z)  \, {\rm Pr}(z|m_0) \\
& \propto & {\mathcal L}(\vec{c} |z)  \, {\rm Pr}(z|m_0). \label{eq:post}
\end{eqnarray} 
Adding the template dependency in the prior would improve our results, but is impractical with the iterative method presented in Sec.~\ref{sec:PDF}, given the size of our sample. 

The posterior probability ${\mathcal P}(z|\obs)$ is a photometric estimate of the true conditional redshift probability $p(z|\obs)$ in Eq.~(\ref{eq:nzpdz}), and thus we are able to estimate the target sample $N(z)$ via stacking of the individual galaxy posterior probability distributions:
\begin{equation} \label{eq:nzpdzpost}
    N(z) = \sum_i^{N_{\rm t}} {\mathcal P}_i(z|\obs),
\end{equation}
and therefore: 
\begin{equation} \label{eq:mzpdz}
    \avz = \frac{\int z \left[ \sum_i^{N_{\rm t}} {\mathcal P}_i(z|\obs) \right] {\rm d} z}{\int \left[ \sum_i^{N_{\rm t}}  \, {\mathcal P}_i(z|\obs) \right] {\rm d} z}.
\end{equation}

\begin{figure*}
\includegraphics[width=18.cm]{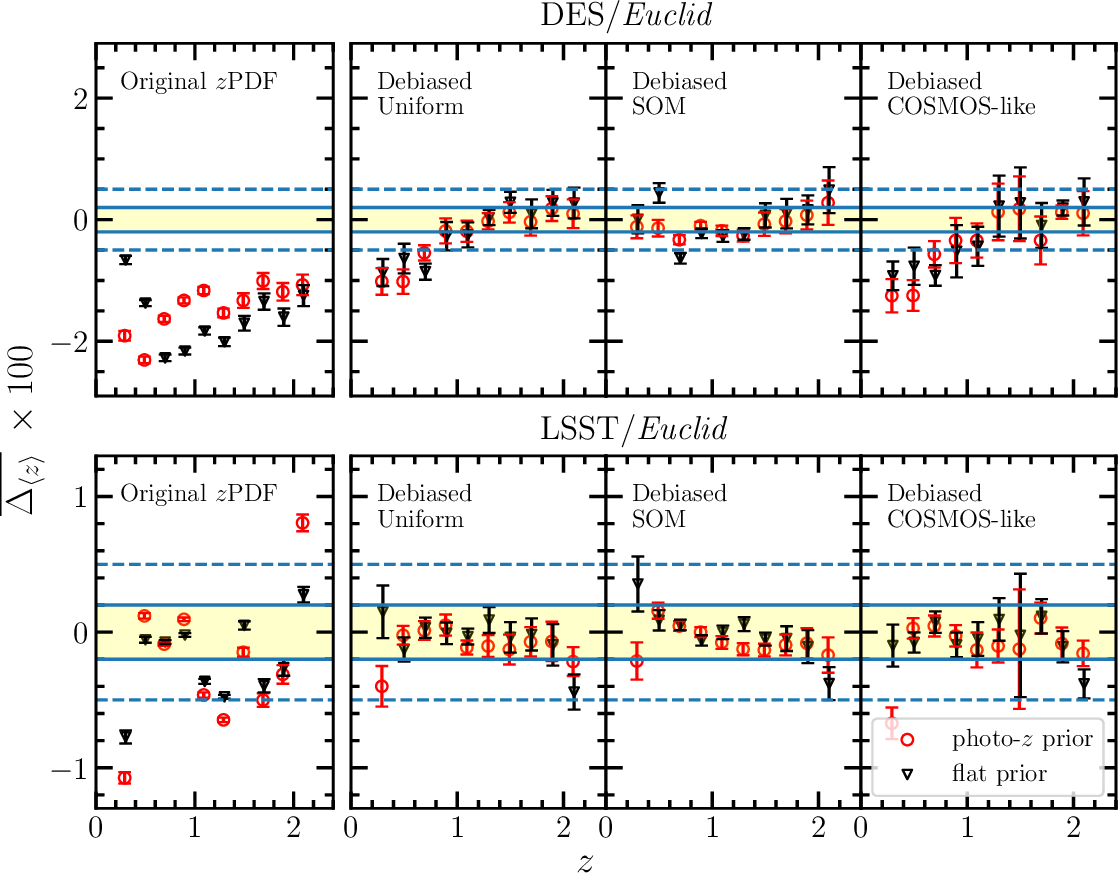}  
\caption{Bias on the mean redshift (see Eq.~\ref{eq:deltaz}), estimated using the \zPDF{} method and averaged over the 18 photometric noise realisations. The top and bottom panels correspond to the DES/\Euclid{} and LSST/\Euclid{} mock catalogues, respectively. Note the differing scales in the $y$-axes of the two panels. The left panels are obtained by summing the initial \zPDF{}, without any attempt at debiasing. The other panels show the results of summing the \zPDF{} after debiasing, assuming (from left to right) a uniform, SOM, and COSMOS-like training sample. The yellow region represents the \Euclid{} requirement of $|\Delta_{\avz}|\,\leq\,0.002\,(1+z)$. The red circles and black triangles in each panel correspond to the results estimated using  \photoz{} and flat priors, respectively.
}
\label{fig:compareNzPrior}
\end{figure*}

\subsection{Initial results}
\label{subsec:simpleSum}

In this analysis we use the \lephare{} code, which outputs ${\mathcal L}(\obs|z)$ for each galaxy as defined in Eq.~(\ref{eq:likedef}). The redshift distribution (and thereafter its mean) are obtained by summing galaxy posterior probabilities, which are derived as in Eq.~(\ref{eq:post}). This raises, however, an immediate concern: in order to estimate the $N(z)$ using the per-galaxy likelihoods, we require a prior distribution of magnitude-dependant redshift probabilities, ${\rm Pr}(z|m_0)$, which naturally requires knowledge of the magnitude-dependent redshift distribution. 

We test the sensitivity of our method to this prior choice by considering priors of two types: a (formally improper) `flat prior' with ${\rm Pr}(z|m_0)=1$; and a `photo-$z$ prior' that is constructed by normalising the redshift distribution, estimated per magnitude bin, as obtained by summation over the likelihoods \citep[following][]{brodwin06}. Formally this photo-z prior is defined as:
\begin{equation}\label{eq:priormag}
    {\rm Pr}(z|m_0) = \sum_i^{N_{\rm t}} {\mathcal L}_i(\obs|z) \,  \Theta(m_{0,i}|m_0),
\end{equation}
where $\Theta(m_{0,i}|m_0)$ is unity if $m_{0,i}$ is inside the magnitude bin centered on $m_0$ and zero otherwise, and $N_{\rm t}$ is the number of galaxies in the tomographic bin. 

We estimate $\avz$ in the previously defined tomographic bins using Eq.~(\ref{eq:mzpdz}). In the upper-left panel of Fig.~\ref{fig:PIT}, we show estimated (and true) $N(z)$ for one tomographic bin with $1.2<z_{\rm p}<1.4$, estimated using DES/\Euclid{} photometry. We annotate this panel with the  estimated $\Delta_{\avz}$ made when utilising our two different priors. It is clear that the choice of prior, in this circumstance, can have a significant impact on the recovered redshift distribution. We also find an offset in the estimated redshift distributions with respect to the truth, as confirmed by the associated mean redshift biases being considerable: $|\Delta_{\avz}|\, > \, 0.012$, or roughly six times larger than the \Euclid{} accuracy requirement.  

The resulting biases estimated for this method in all tomographic bins, averaged over all noise realisations, is presented in the left-most panels of Fig.~\ref{fig:compareNzPrior} (for both the DES/\Euclid{} and LSST/\Euclid{} configurations). Overall, we find that this approach produces mean  biases of $|\biaszmeanB{}|\,>\,0.02\,(1+z)$ and $|\biaszmeanB{}|\,>\,0.01\,(1+z)$, which corresponds to roughly ten and five times larger than the \Euclid{} accuracy requirement, for the DES/\Euclid{} and LSST/\Euclid{} cases respectively. Such bias is created by the mismatch between the simple galaxy templates included in \lephare{} (in a broad sense, including dust attenuation and IGM absorption) and the complexity and diversity of galaxy spectra generated in the hydrodynamical simulation. Such biases are in agreement with the usual values observed in the literature with broad band data \citep[e.g.][]{hildebrandt12}.

We therefore conclude that use of such a redshift calibration method is not feasible for \Euclid{}, even under optimistic photometric circumstances. 

\subsection{Redshift probability debiasing}
\label{subsec:BordoMethod}

In the previous section we demonstrated that the estimation of galaxy redshift distributions via summation of individual galaxy posteriors ${\mathcal P}(z)$, estimated with a standard template-fitting code, is too inaccurate for the requirements of the \Euclid{} survey. The cause of this inaccuracy can be traced to a number of origins: colour-redshift degeneracies, template set non-representativeness, redshift prior inadequacy, and more. However, it is possible to alleviate some of this bias, statistically, by incorporating additional information from a spectroscopic training sample. In particular, \citet{bordoloi10} proposed a method to debias ${\mathcal P}(z)$ distributions, using the Probability Integral Transform \citep[PIT, ][]{Dawid84}. The PIT of a distribution is defined as the value of the cumulative distribution function evaluated at the ground truth. In the case of redshift calibration, the PIT per galaxy is therefore the value of the cumulative ${\mathcal P}(z)$ distribution evaluated at source spectroscopic redshift $z_{{\rm s}}$:
\begin{equation}
    {\rm PIT} = {\mathcal C}(z_{{\rm s}}) = \int^{z_{{\rm s}}}_0 {\mathcal P}(z) \, {\rm d} z.
\end{equation}
If all the individual galaxy redshift probability distributions are accurate, the PIT values for all galaxies should be uniformly distributed between 0 and 1. Therefore, using a spectroscopic training sample, any deviation from uniformity in the PIT distribution can be interpreted as an indication of bias in individual estimates of ${\mathcal P}(z)$ per galaxy. We define $N_{\rm P}$ as the PIT distribution for all the galaxies within the training spectroscopic sample, in a given tomographic bin. \citet{bordoloi10} demonstrate that the individual ${\mathcal P}(z)$ can be debiased using the $N_{\rm P}$ as:
\begin{equation}\label{eq:debias}
    {\mathcal P}_{\rm deb}(z) =  {\mathcal P}(z) \times N_{\rm P}[{\mathcal C}(z)] \; \left[ \; \int^{1}_{0} N_{\rm P}(x) \, {\rm d} x \right]^{-1},
\end{equation} 
where ${\mathcal P}_{\rm deb}(z)$ is the debiased posterior probability, and the last term ensures correct normalisation. This correction is performed per tomographic bin. 

This method assumes that the correction derived from the training sample can be applied to all galaxies of the target sample. As with the direct calibration method, such an assumption is valid only if the training sample is representative of the target sample, i.e. in the case of a uniform training sample, but not in the case of the COSMOS-like and SOM training samples. In these latter cases, we weight each galaxy of the training sample in a manner equivalent to the direct calibration method (see Sect.~\ref{sec:direct}), in order to ensure that the PIT distribution of the training sample matches that of the target sample (which is of course unknown). As for direct calibration, a completely missing population (in redshift or spectral type) could impact the results in an unknown manner, but such case should not occur for a uniform or SOM training sample.

Until now we have considered two types of redshift prior (defined in Sect.~\ref{subsec:simpleSum}): (1) the flat prior and (2) the \photoz{} prior. We have shown that the choice of prior can have a significant impact on the recovered $\avz$ (Sect.~\ref{subsec:simpleSum}). However, as already noted by \citet{bordoloi10}, the PIT correction has the potential to account for the redshift prior implicitly. In particular, if one uses a flat redshift prior, the correction essentially modifies ${\mathcal L}(z)$ to match the true ${\mathcal P}(z)$ (assuming the various assumptions stated previously are satisfied). This is because the redshift prior information is already contained within the training spectroscopic sample. 
Nonetheless, rather than assuming a flat prior to measure the PIT distribution, one can also adopt the \photoz{} prior (as in Eq.~\ref{eq:priormag}). This approach has two advantages: (1) it allows us to start with a posterior probability that is intrinsically closer to the truth, and (2) it includes the magnitude dependence of the redshift distribution within the prior, which is of course not reflected in the case of the flat prior. 

Therefore, we improve the debiasing procedure from \citet[][]{bordoloi10} by including such \photoz{} prior. We add an iterative process to further ensure the correction's fidelity and stability. In this process the PIT distribution is iteratively recomputed by updating the \photoz{} prior. We compute the PIT for the galaxy as:
\begin{equation}
{\mathcal C}^{n}(z_{\rm s}) \,  = \, \int^{z_{\rm s}}_0 {\mathcal L}(z) \, {\rm Pr}^n(z|m_0) \, {\rm d}z,
\end{equation}
where ${\rm Pr}^n(z|m_0)$ is the prior computed at step $n$.
We can then derive the debiased posterior as:
\begin{equation}
    {\mathcal P}^n_{\rm deb}(z) \, = \,  {\mathcal L}(z) \, {\rm Pr}^n(z|m_0) \; \times \; N_{\rm P}^n [ {\mathcal C}^{n}(z) ],
\end{equation}
with $N_{\rm P}^n$ the PIT distribution at step $n$. The prior at the next step is:
\begin{equation}
    {\rm Pr}^{n+1}(z|m_0) \, = \, \sum_i^{N_{\rm T}} {\mathcal P}^n_{{\rm deb}, i}(z|\obs) \, \Theta(m_i|m_0),
\end{equation}
with $m_i$ for the magnitude of the galaxy $i$. 
Note that at $n=0$, we assume a flat prior. Therefore, the step $n=0$ of the iteration corresponds to the debiasing assuming a flat prior, as in \citet[][]{bordoloi10}. We also note that the prior is computed for the $N_{\rm T}$ galaxies of the training sample in the debiasing procedure, while it is computed over all galaxies of the tomographic bin for the final posterior.

As an illustration, Fig.~\ref{fig:PDFexamples} shows the debiased posterior distributions with black lines, which can significantly differ from the original likelihood distribution.
We find that this procedure converges quickly. Typically, the difference between the mean redshift measured at step $n+1$ and that measured at step $n$ does not differ by more than $10^{-3}$ after 2--3 iterations. 

As described in appendix \ref{idealized}, we also find that the debiasing procedure is considerably more accurate when the \photoz{} uncertainties are over-estimated, rather than under-estimated. Such a condition can be enforced for all galaxies by artificially inflating the source photometric uncertainties by a constant factor in the input catalogue, prior to the measurement of \photoz{}. In our analysis, we utilise a factor of two inflation in our photometric uncertainties prior to measurement of our \photoz{} in our debiasing technique.

\subsection{Final results}
\label{subsec:BordoApply}

We illustrate the impact of the ${\mathcal P}(z)$ debiasing on the recovered redshift distribution in the lower panels of Fig.~\ref{fig:PIT}. This figure presents the case of the redshift bin $0.8<z_{\rm p}<1$ in the DES/\Euclid{} configuration. The $N(z)$ and PIT distributions, as computed with the initial posterior distribution are shown in the upper panels (for both of our assumed priors). The distributions after debiasing are shown in the bottom panels. We can see the clear improvement provided by the debiasing procedure in this example, whereby the redshift distribution bias $\Delta_{\avz}$ (annotated) is reduced by a factor of ten. We also observe a clear flattening of the target sample PIT distribution.

We present the results of debiasing on the mean redshift estimation for all tomographic bins in Fig.~\ref{fig:compareNzPrior}. The three right-most panels show the mean redshift biases recovered by our debiasing method, averaged over the 18 photometric noise realisations, for our three training samples. The accuracy of the mean redshift recovery is systematically improved compared to the case without ${\mathcal P}(z)$ debiasing (shown in the left column). In the DES/\Euclid{} configuration for instance (shown in the upper row), the improvement is better than a factor of ten at $z>1$. In the LSST/\Euclid{} configuration (shown in the bottom row), we find that the results do not depend strongly on the training set used: the accuracy of $\avz$ is similar for the three training samples, showing that stringent control of the representativeness of the training sample is not necessary in this case. In the DES/\Euclid{} case, however, the SOM training sample clearly out-performs the other training samples, especially at low redshifts. Finally, we note that the iterative procedure using the \photoz{} prior improves the results when using the SOM training sample and the DES/\Euclid{} configuration.

Overall, the \Euclid{} requirement on redshift calibration accuracy is not reached by our debiasing calibration method in the DES/\Euclid{} configuration. The values of \biaszmean{} at $z<1$ reach five times the \Euclid{} requirement, represented by the yellow bands in Fig.~\ref{fig:compareNzPrior}. At best, an accuracy of $|\biaszmeanB{}|\,\leq\,0.004\,(1+z)$ is reached for the SOM training sample with the \photoz{} prior. Conversely, the \Euclid{} requirement is largely satisfied in the LSST/\Euclid{} configuration. In this case, biases of $|\biaszmeanB{}|\,\leq\, 0.002\,(1+z)$ are observed in all but the two most extreme tomographic bins: $0.2<z<0.4$ and $2<z<2.2$. We therefore conclude that, for this approach, deep imaging data is crucial to reach the required accuracy on mean redshift estimates for \Euclid{}. 

\section{Discussion on key model assumptions}
\label{sec:discussion} 

In this section, we discuss how some important parameters or assumptions impact our results. We start by discussing the impact of catastrophic redshift failures in the training sample, the impact of our pre-selection on photometric redshift uncertainty, and the influence of the size of the training sample on our conclusions. We also discuss some remaining limitations of our simulation in the last subsection.

\begin{figure}
\includegraphics[width=0.96\columnwidth]{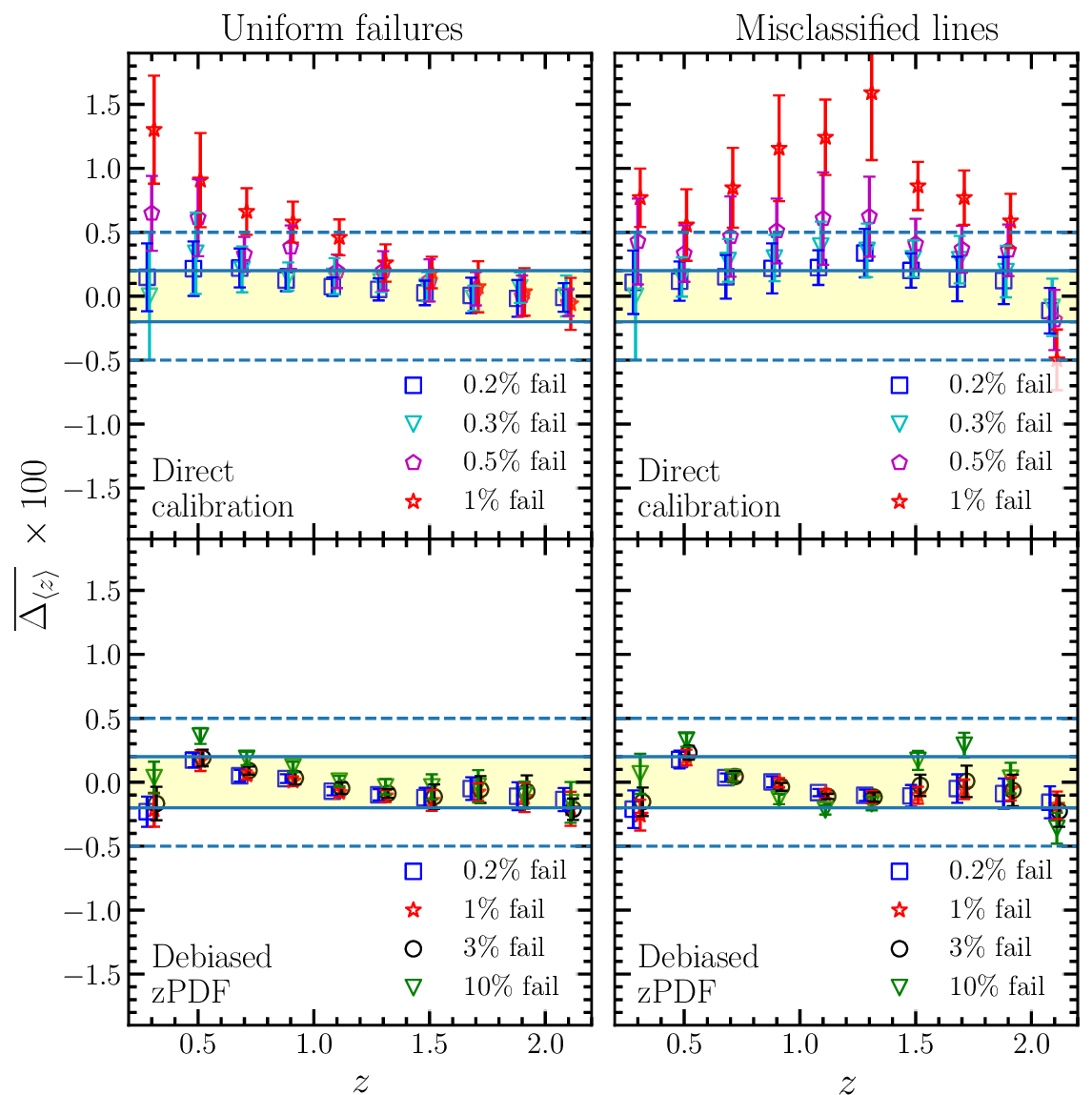} 
\caption{Bias on the mean redshift averaged over the 18 photometric noise realisations in the LSST/\Euclid{} case. We assume a SOM training sample, and the different symbols correspond to various fraction of failures introduced in the \specz{} training sample. The left and right panels correspond to different assumptions on how to distribute the catastrophic failures in the \specz{} measurements: uniformly distributed between $0<z<4$ (left), and assuming failures are caused by misclassified emission lines (right). The upper and lower panels correspond to the direct calibration and debiasing method, respectively.
\label{fig:specErr}}
\end{figure}

\subsection{Impact of catastrophic redshift failures in the training sample}
\label{subsec:failures}

For all results presented in this work so far, we have assumed that spectroscopic redshifts perfectly recover the true redshift of all training sample sources. However, given the stringent limit on the mean redshift accuracy in \Euclid{}, deviations from this assumption may introduce significant biases. In particular, mean redshift estimates are extremely sensitive to redshifts far from the main mode of the distribution, and therefore catastrophic redshift failures in spectroscopy may present a particularly significant problem. For instance, if $0.5\%$ of a galaxy population with true redshift of $z=1$ are erroneously assigned $z_{\rm s}>2$, then this population will exhibit a mean redshift bias of $|\biaszmeanB{}|>0.002$ under direct calibration.

Studies of duplicated spectroscopic observations in deep surveys have shown that there exists, typically, a few percent of sources that are assigned both erroneous redshifts and high confidences \citep[e.g.][]{lefevre05}. Such redshift measurement failures can be due to misidentification between emission lines, incorrect associations between spectra and sources in photometric catalogues, and/or incorrect associations between spectral features and galaxies \citep[due, for example, to the blending of galaxy spectra along the line of sight;][]{masters17,urrutia19}. Of course, the fraction of redshift measurement failures is dependant on the observational strategy (e.g. spectral resolution) and the measurement technique (e.g. the number of reviewers per observed spectrum). Incorrect association of stars and galaxies can also create difficulties. Furthermore, the frequency of redshift measurement failures is expected to increase as a function of source apparent magnitude; a particular problem for the faint sources probed by \Euclid{} imaging (\RIZ$<24.5$). 

As we cannot know a priori the number (nor location) of catastrophic redshift failures in a real spectroscopic training set, we instead estimate the sensitivity of our results to a range of catastrophic failure fractions and modes. We assume a SOM-based training sample and an LSST/\Euclid{} photometric configuration, and distribute various fractions of spectroscopic failures throughout the training sample, simulating both random and systematic failures. Generally though, because these failures occur in the spectroscopic space, recovered calibration biases are largely independent of the depth of the imaging survey and the method used to build the training sample. 

We start by testing the simplest possible mechanism of distributing the failed redshifts, by assigning failed redshifts uniformly within the interval $0<z<4$. Resulting calibration biases for this mode of catastrophic redshift failure are presented in the left panels of Fig.~\ref{fig:specErr}. We find that, for the direct calibration approach (top panel), even 0.2\% of failures in the training sample is the limit to bias the mean redshift by $|\mu_{\Delta z}|\, > \, 0.002$ at low redshifts \citep[by definition, flag 3 in the VVDS could include 3\% of failures;][]{lefevre05}. We also find that the bias decreases with redshift and reaches zero at $z=2$. This is a statistical effect; our assumed uniform distribution has a $z=2$ mean, and so random catastrophic failures scattered about this point induce no shift in a $z\approx2$ tomographic bin. For the same reason, biases would be significant at the two extreme tomographic bins if we were to assume a catastrophic failure distribution that followed the true $N(z)$ (which peaks at $z\approx1$). In contrast, our debiased \zPDF{} approach is found to be resilient to catastrophic failure fractions as high as 3.0\% (bottom panel). In that case, only an unlikely failure fraction of 10\% biases the mean redshift by $|\biaszmeanB{}|\,\geq\,0.002\,(1+z)$. We interpret this result demonstrating the low sensitivity of the PIT distribution to redshift failures in the training sample. This is related to the fact that the PIT distribution provides a global statistical correction that is only weakly sensitive to individual galaxy redshifts.

In the previous test, we assign the failed redshifts uniformly within the interval $0<z<4$, which is not the expected distribution when redshift failures occur by misidentification of spectral emission lines \citep[e.g.][]{lefevre15, urrutia19}. This mode of failure leads to a highly non-uniform distribution of failed redshifts, due to the interplay between the location of spectral emission lines and the redshift distribution of training sample galaxies. If a line emitted at $\lambda_{\rm true}$ is misclassified as a different emission line at $\lambda_{\rm wrong}$, the redshift is therefore assigned to be:  
\begin{equation}
z_{\rm wrong} = \frac{\lambda_{\rm true}}{\lambda_{\rm wrong}}(1+z_{\rm true})-1.
\end{equation}
We study the impact of such line misidentifications on our estimates of $\avz$, by introducing redshift failures in the simulation with the following assumptions:
\begin{itemize}
\item  if $z_{\rm true}<0.5$, we assume that the H$_\alpha$ emission line can be misclassified as [O{\sc ii}]; 
\item  if $0.5<z_{\rm true}<1.4$, we assume that [O{\sc ii}] can be misclassified as H$_\alpha$ (for bright sources) or Ly$_\alpha$ (for faint sources, using $i=23.5$ as a limit); 
\item at $1.4<z_{\rm true}<2.0$, we assume that the redshift is estimated using NIR spectra, and therefore that the H$_\alpha$ line can be misclassified as [O{\sc ii}]; 
\item and for sources at $z>2$, we assume that Ly$_\alpha$ can be misclassified as [O{\sc ii}]. 
\end{itemize}
The same fraction of misclassifications is assumed in all the redshift intervals. The result of this experiment is shown in the right panels of Fig.~\ref{fig:specErr}, and demonstrates that this (more realistic) mode of catastrophic failures results in equivalent levels of bias as was seen in our simple (uniform) mode, albeit in different tomographic bins. This confirms that the sensitivity of the direct calibration to catastrophic redshift failures exists across simplistic and complex failure modes. In this mode, a failure fraction of 0.2\% is sufficient to bias direct calibration at $|\biaszmeanB{}|\,\geq\,0.002\,(1+z)$ in all tomographic bins with $z_{\rm p}>0.6$. This highlights that the calibration bias depends on the exact distribution of failed redshifts: in the case of line misidentification, incorrectly assigned redshifts consistently bias spectra to higher redshift, causing $\avz{}$ to be affected more heavily over the full redshift range. 

We compare our result to the simulation of \citet{wright20}. They investigate the impact of catastrophic \specz{} failures on the estimate of $\avz$ (for KiDS cosmic shear analyses) in the MICE2 simulation \citep{Fosalba15}. They introduce $1.03\%$ of failed redshifts following various distributions. In particular, they test the case of a uniform distribution within $0<z<1.4$, where $z=1.4$ is the limiting redshift of the MICE2 simulation. They report a bias in their direct calibration of $\Delta_{\avz}=0.0029$ for their lowest redshift tomographic bin, and smaller biases for higher redshift tomographic bins. In our lowest redshift bin, we observe a bias of $\Delta_{\avz}=0.01$ for a similar analysis. We argue that this is entirely consistent with the results of \citet{wright20} given that our considered redshift range is almost three times larger. \citet{wright20} conclude that \specz{} failures are unlikely to influence cosmic shear analyses with the KiDS survey, which are limited to $z<1.2$, but may be significant for \Euclid{}-like analyses. In this way, our results also agree; it is clear that direct calibration for next generation (so called `Stage-IV') cosmic-shear surveys like \Euclid{} will require careful consideration of the influence of catastrophic spectroscopic failures. 

The training sample for \Euclid{} is currently being built with the  C3R2 survey \citep{masters19,guglielmo20_c3r2}. Such sample results from a combination of spectra coming from numerous instruments installed on 8-meter class telescopes (e.g. VIMOS, FORS2, KMOS, DEIMOS, LRIS, MOSFIRE) including data from previous spectroscopic surveys \citep[e.g.][]{Lilly:2007p5770,lefevre15,kashino19}. The most robust \specz{} acquired on the \Euclid{} Deep fields with the NISP instrument will be included. Given the diversity of observations, a careful assessment of the sample purity is necessary to limit the fraction of failures below 0.2\%.
Encouragingly, \citet{masters19} do not find any redshift failures within the 72 C3R2 \specz{} with duplicated observations. Nonetheless, a larger sample of confirmed spectra is necessary to demonstrate that less than 0.2\% of spectroscopic redshift measurements suffer from catastrophic failure. Finally, it is possible that improved reliability of both direct calibration methods and spectroscopic confidence could decrease the effects seen here: \citet{wright20}, for example,  advocate a means of cleaning cosmic shear photometric samples of sources with poorly constrained mean redshifts, demonstrating that this can cause a considerable reduction in calibration biases. Of course, the problem could possibly be alleviated if one were able to improve the reliability of the training sample by only including \specz{} with corroborative evidence from, for example, high-precision \photoz{} derived from deep photometry in the calibration fields.

\begin{figure}
    \centering
    \includegraphics[width=0.96\columnwidth]{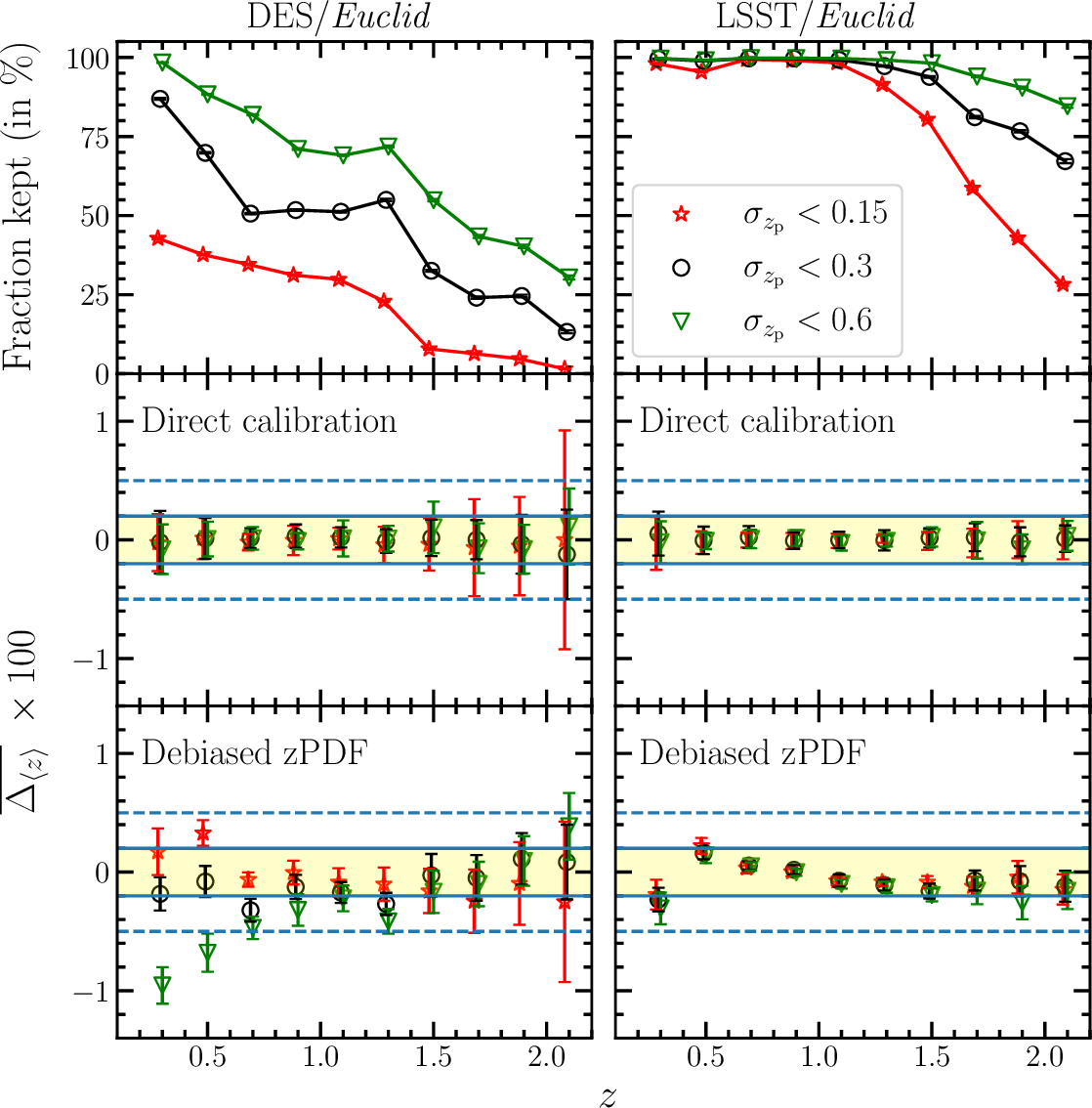}
    \caption{Bias on the mean redshift (see Eq.~\ref{eq:deltaz}), averaged over the 18 photometric noise realisations, under different $\sigma_{z_{\rm p}}$ selection thresholds.
    Top panels: fraction of the sample retained after having applied different $\sigma_{z_{\rm p}}$  thresholds. The middle and bottom panels show the bias on the mean redshift using the direct calibration and debiasing technique, respectively. The left and right panels correspond to the DES/\Euclid{} and LSST/\Euclid{} configurations, respectively. We assume a SOM training sample with 2 gal/cell.}
    \label{fig:testErr}
\end{figure} 

\subsection{Relaxing the \photoz{} $\sigma_{z_{\rm p}}$ preselection}
\label{subsec:discussionErr} 

Estimates of the redshift distribution mean are also sensitive to the presence of secondary modes in the redshift distribution, and our ability to reconstruct them. As described in Sect.~\ref{subsec:photoz}, all results presented thus far have invoked a selection on the photometric redshift uncertainty of $\sigma_{z_{\rm p}}<0.3$, which reduces the likelihood of secondary redshift distribution peaks in our analysis. Here we discuss the impact of this adopted threshold on both accuracy of our estimates of $\avz$, and on the fraction of photometric sources that satisfies this selection (and so are retained for subsequent cosmic shear analysis). We apply several $\sigma_{z_{\rm p}}$ thresholds in the range $\sigma_{z_{\rm p}}\,\in\,[0.15,0.6]$ to the full \photoz{} catalogue. For the training sample, we consider the SOM configuration with two galaxies per cell. The results are shown in Fig.~\ref{fig:testErr} for the DES/\Euclid{} (left) and LSST/\Euclid{} (right) configurations. We find that the $\sigma_{z_{\rm p}}$ threshold does not influence our conclusions regarding the direct calibration approach, which is largely insensitive to variations in this threshold. We note, however, that the scatter on the mean redshift (\scatterzmean, shown by the errorbars) increases well above the \Euclid{} requirement (for the DES/\Euclid{} configuration) when selecting \photoz{} with $\sigma_{z_{\rm p}}<0.15$; however this is primarily because such a selection drastically reduces the size of the training sample at $z>1.2$, increasing the influence of Poisson noise. Therefore, given the insensitivity of the direct calibration to this threshold, it is advantageous to keep galaxies with broad redshift likelihoods in the target sample when using this method. Conversely, $\sigma_{z_{\rm p}}$ has a decisive impact on the accuracy of mean redshift estimates inferred from the debiased \zPDF{} approach. For instance, in the DES/\Euclid{} configuration, $|\biaszmeanB{}|$ is strongly degraded when applying a threshold of $\sigma_{z_{\rm p}}<0.6$. Such a threshold on $\sigma_{z_{\rm p}}$ could be relaxed in the   LSST/\Euclid{} configuration, however, primarily because the sample is already dominated by galaxies with a narrow \zPDF{}.

Not considered in the above, however, is the importance that the target sample number density plays in cosmic shear analyses. Cosmological constraints from cosmic shear are approximately proportional to the square root of the size of the target galaxy sample, and to the mean redshift. Therefore, optimal lensing surveys require a sufficiently high surface density of sources, preferentially at high redshift. In the \Euclid{} project, 30 galaxies per arcmin$^2$ are required to reach their planned scientific objectives \citep{laureijs11}. As shown in the top panels of Fig.~\ref{fig:testErr}, however, applying a threshold on $\sigma_{z_{\rm p}}$ naturally introduces a reduction in the size of the target sample. For instance, we keep less than 10\% of the galaxies at $z>1.4$ by selecting a sample at $\sigma_{z_{\rm p}}<0.15$ in the DES/\Euclid{} configuration. In the LSST/\Euclid{} case, a threshold of $\sigma_{z_{\rm p}}<0.3$ has only a significant impact in the redshift bins above $z>1.6$. A compromise is therefore needed between the number of sources retained in the target sample, and the accuracy of the mean redshift that we estimate for these sources (when using the debiasing technique). We do not attempt to estimate what this optimal selection is using our simulations, as the luminosity function predicted by Horizon-AGN does not perfectly reproduce what is found in real data. Nonetheless, we note that the fraction of galaxies that are removed from the target sample is likely overestimated here: modern cosmic shear analyses typically introduce a weight associated with the accuracy of each source's shape measurement (the `shear weight', which is not included in our simulations), which systematically decreases the contribution of low signal-to-noise galaxies to the analysis. As these fainter sources have intrinsically broader \photoz{} distributions, they will be the most heavily affected by our cuts on $\sigma_{z_{\rm p}}$. 

\begin{figure}
    \centering
    \includegraphics[width=0.96\columnwidth]{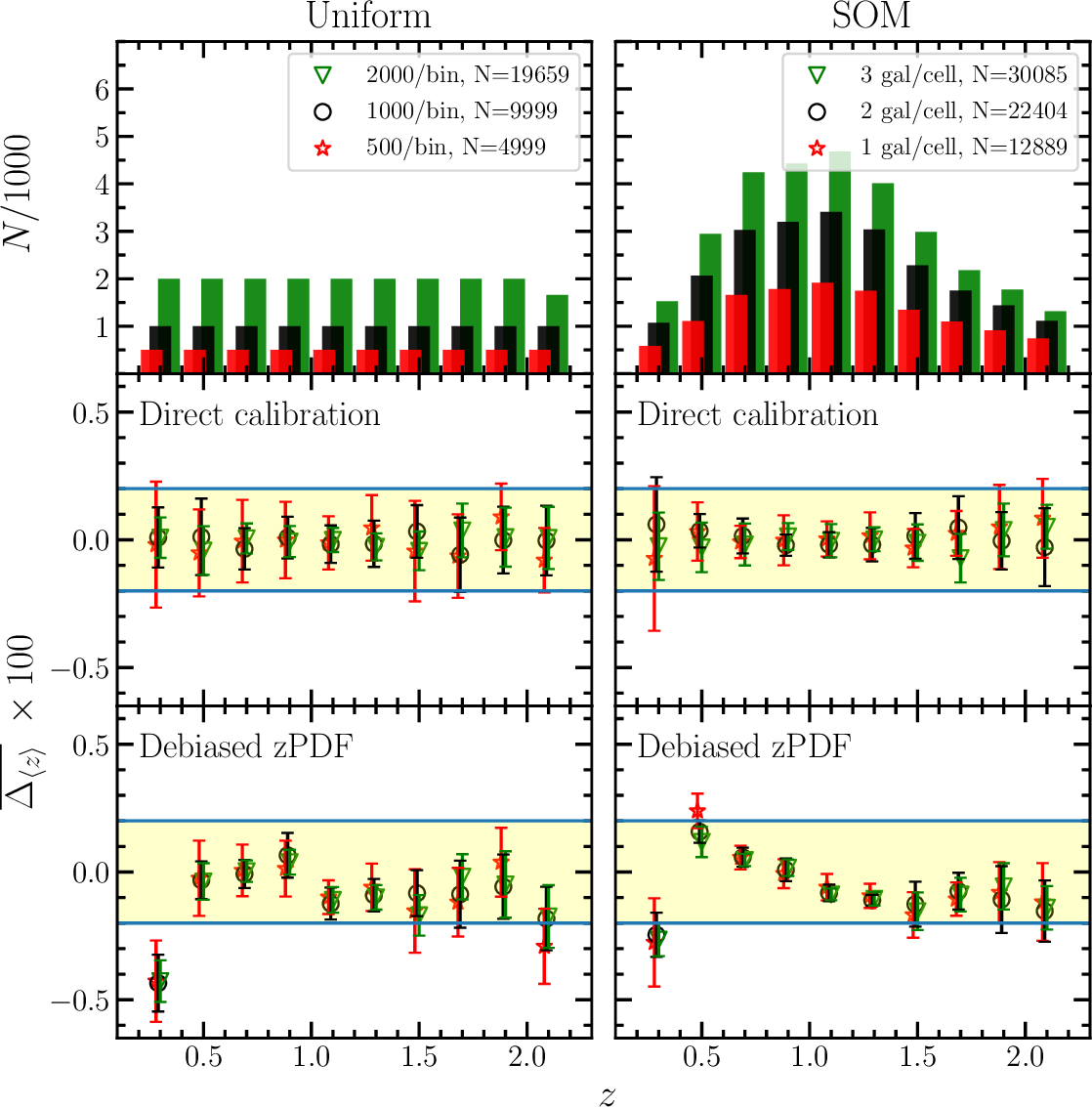}
    \caption{Bias on the mean redshift (see Eq.~\ref{eq:deltaz}) averaged over the 18 photometric noise realisations. Impact of the training sample size on the mean redshift accuracy in the LSST/\Euclid{} case. Left and right panels correspond to a uniform and SOM spectroscopic coverage, respectively. The top panels show the number of galaxies used for the training in three considered cases. Middle and bottom panels show the mean redshift accuracy using the direct calibration and the optimised \zPDF{}, respectively.}
    \label{fig:compareSize}
\end{figure} 

\subsection{Size of the training sample}
\label{subsec:size} 

The size of the training sample is naturally of most importance when using the direct calibration approach \citep[e.g.][]{newman08}. The debiased \zPDF{} approach, though, is also sensitive to statistical noise in the PIT distribution. As some ongoing spectroscopic surveys are designed to produce the training samples for Stage IV weak-lensing experiments \citep[e.g.][]{masters17}, we explore here the minimal size of these samples required for accurate redshift calibration. To do this, we modify the size of the training samples (limiting our analysis to the uniform and SOM training sample cases). We do not consider the COSMOS-like case that is a patchwork of existing surveys, and is not specifically designed for weak-lensing experiments. For the uniform training samples, we test the cases with 500, 1000, 2000 galaxies per tomographic bin. For the SOM training samples, we test the cases corresponding to cells filled with 1, 2, or 3 galaxies. 

Figure~\ref{fig:compareSize} shows the impact of the training sample size on $\Delta_{\avz}$. We find that the mean bias \biaszmean{} always remains within the \Euclid{} requirements for the direct calibration approach. The scatter \scatterzmean{} in the bias exceeds the \Euclid{} requirements in few tomographic bins, however only when considering the smallest training samples: the \Euclid{} requirements are fully satisfied in all tomographic bins when assuming a training sample with more than 1000 galaxies per bin or more than two galaxies per SOM cell. With the debiased \zPDF{} approach, we find that increasing the size of the training sample is not sufficient to reduce the residual bias in the method; rather deeper photometry is preferable, to improve the quality of the initial \zPDF{}.

\subsection{Catastrophic failures within the \photoz{} sample}
\label{subsec:cata} 

Catastrophic failures in the \photoz{} sample are a concern for both methods described in this paper. We discuss here their impact as well as remaining limitations of our simulation.

As shown in Fig.~\ref{fig:zmlzs}, our simulated sample already includes a significant fraction of \photoz{} outliers, defined such that $|z_{\rm p}-z_{\rm s}|>0.15\,\left(1+z_{\rm s}\right)$. We find 16.24\% and 0.70\% of outliers at $\RIZ{}<24.5$ in DES/\Euclid{} and LSST/\Euclid{}, respectively. These fractions reduce to 1.82\% and 0.04\% when applying a selection on the photometric redshift uncertainty at $\sigma_{z_{\rm p}}<0.3$. The largest fraction of these outliers is due to the degeneracies in the colour-redshift space inherent to the use of low signal-to-noise photometry in several bands. However, less trivial catastrophic failures are also present in the simulation. In particular, the diversity of spectra generated by the complex physical processes in Horizon-AGN is not fully captured by the limited set of SED templates used in \lephare{}. This misrepresentation in galaxy SED creates a significant fraction of \zPDF{} not compatible with the \specz{}. An example of such \like{} is shown in the bottom right panel of Fig.~\ref{fig:PDFexamples}. Despite the presence of such failures, our results show that the \Euclid{} requirement is fulfilled.

Several factors were ignored that can potentially create more catastrophic failures in the \photoz{}. Galaxies with extreme properties, such as sub-millimeter galaxies (SMG) for instance, are known to be under-represented in simulations \citep[e.g.][]{hayward20_smg}. If galaxies with an extreme dust attenuation fall within the cosmic-shear selection at $\RIZ{}<24.5$ and are selected in one tomographic bin, they could have an impact on our results. Nonetheless, nothing indicates that their \zPDF{} cannot be established correctly from template fitting, or that such population cannot be isolated in the multi-color space with SOM. 

The presence of AGN could also be a problem. These sources can be isolated from their SED \citep[][]{fotopoulo18}, identified as point-like sources for quasi-stellar objects, and identified as X-ray sources with eROSITA \citep[][]{merloni12}. We should however fail to isolate AGN with an extended morphology or that are too faint to be detected in X-ray. \citet[][]{salvato11} find however that standard galaxy SED libraries are sufficient to obtain an accurate \photoz{} for such sources. 

Residual contamination from stars could also bias $\avz{}$. This population contaminates preferentially specific tomographic bins. In particular, stars may bias the mean redshift towards higher values, for both direct calibration and debiased \zPDF{} methods. A morphological selection based on VIS high-resolution images, combined with a color selection including near-infrared photometry \citep[e.g.][]{daddi04}, is efficient to isolate them \citep[][]{fotopoulo18}. A minimal contamination could bias the mean redshift at a level similar to the one discussed in Sect.\ref{subsec:failures}. Nonetheless, future simulations need to include stellar and AGN populations to better assess the level of contamination of the galaxy sample and its impact on the \Euclid{} requirement.

Finally, \citet[][]{laigle19} show that the fraction of outliers in Horizon-AGN remains underestimated in comparison to real dataset. One source of discrepancy originates from not taking into account the uncertainties induced by source extraction in images. \citet[][]{bordoloi10} estimate that 10\% of the sources could be potentially blended and that the likelihood of two blended galaxies with a magnitude difference lower than two is affected in an unpredictable way. In the last decade, numerous source extraction methods have been developed to perform photometry in crowded fields \citep{desantis07_convphot, laidler07_tfit, merlin16_tphot,lang_tractor_2016}, which could mitigate the impact of blending. Therefore, a new set of simulations that include images and such source extraction tools should be considered in the future.

\section{Application to real data}
\label{sec:data}

In this section, we apply the two approaches presented in Sect.~\ref{sec:direct} and Sect.~\ref{sec:PDF} to real data. We use existing imaging surveys and associated \photoz{} to define several tomographic bins. In each tomographic bin, we select a subsample of \specz{} for which the mean redshift $\avz_{\rm true}$ is known. We refer to this sample as the target sample and the goal is to retrieve the mean redshift  
using only the photometric catalogue and an independent training sample. As previously, we measure $\Delta_{\avz}$ as defined in Eq.~(\ref{eq:deltaz}) in each tomographic bin.

\subsection{The COSMOS survey}
\label{subsec:COSMOS-LEGAC}

We first investigate a favourable configuration, where the photometric survey is much deeper than the target sample. We aim at measuring the mean redshift of the LEGA-C galaxies \citep{vanderwelt16} selected in the tomographic bin at $0.7<z_{\rm p}<0.9$. We base our estimate of $\avz{}$ on the COSMOS broad-band photometry and associated \zPDF{}. The imaging sensitivity is three magnitudes deeper than that of the target sample. All the \specz{} available on the COSMOS field (excluding the LEGA-C ones) are used for the training. For the direct calibration approach, we obtain a bias of $\biaszmeanB{}=0.00032$ and a scatter of $\sigma_{\Delta z}=0.00135$; an accuracy well within the \Euclid{} requirement. Secondly, we debias the \zPDF{} using the PIT distribution as discussed in Sect.~\ref{subsec:BordoMethod}. In that case, we obtain a mean redshift with a bias of $\biaszmeanB{}=-0.00046$ and a scatter of $\sigma_{\Delta z}=0.00073$. In the case of a target sample associated with much deeper photometry, we thus reach the $0.002\,(1+z)$ accuracy requirement of \Euclid{}, either using the direct calibration or debiased \zPDF{} approaches. The details of this measurement are given in Appendix \ref{appendix:COSMOS-LEGAC}.

\begin{figure*}
\begin{tabular}{l l}
\includegraphics[width=8.5cm]{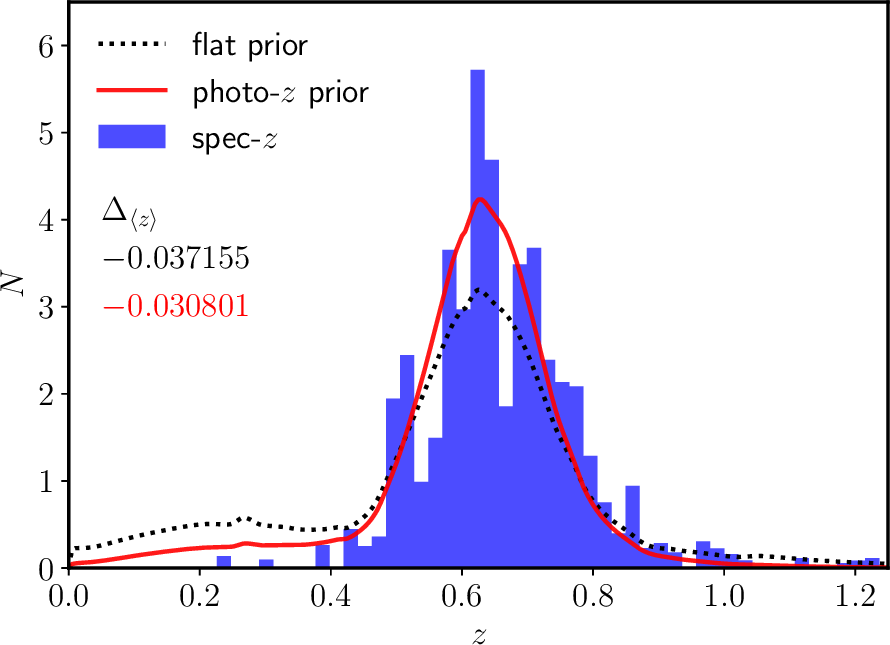}  &
\includegraphics[width=8.5cm]{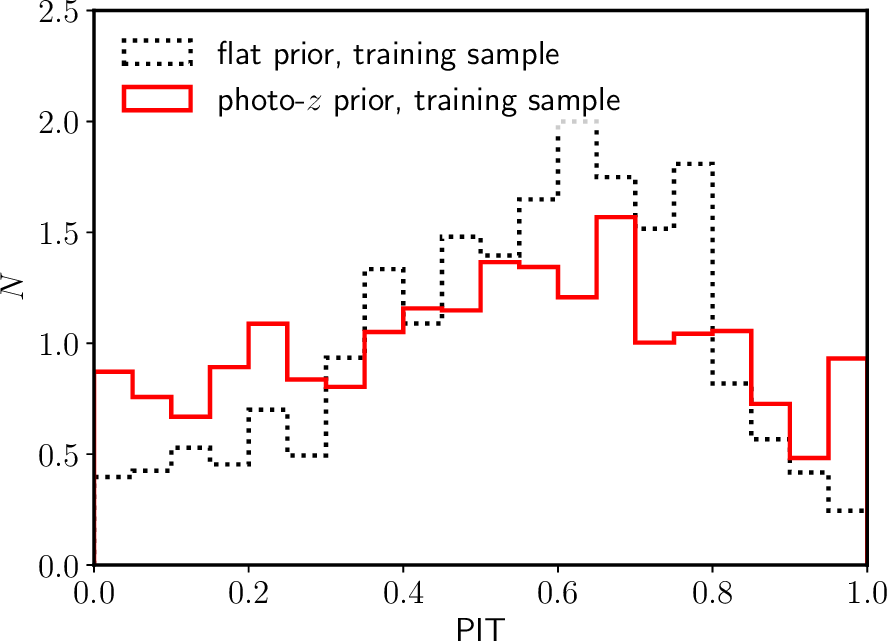} \\   
\includegraphics[width=8.5cm]{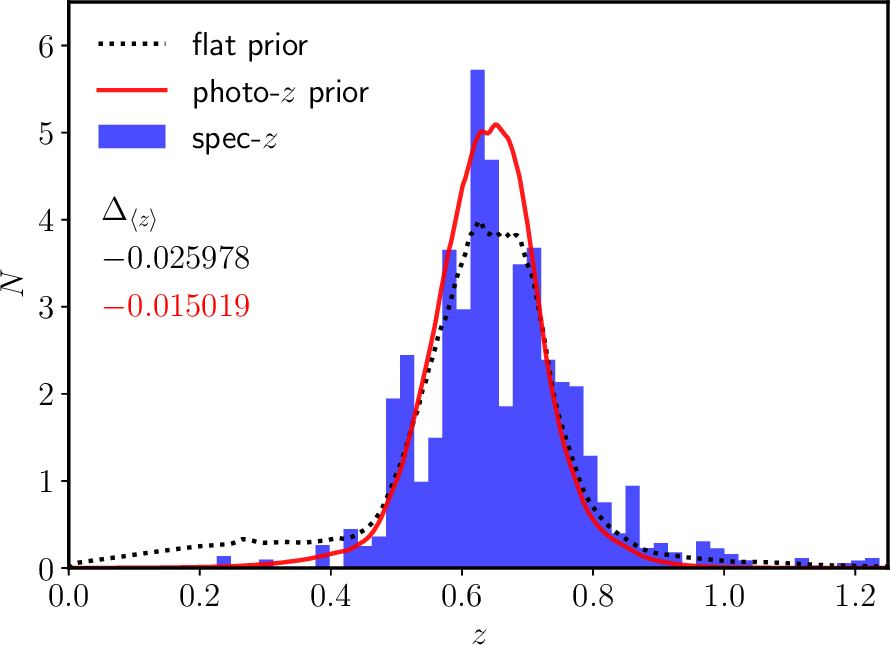}  &
\includegraphics[width=8.5cm]{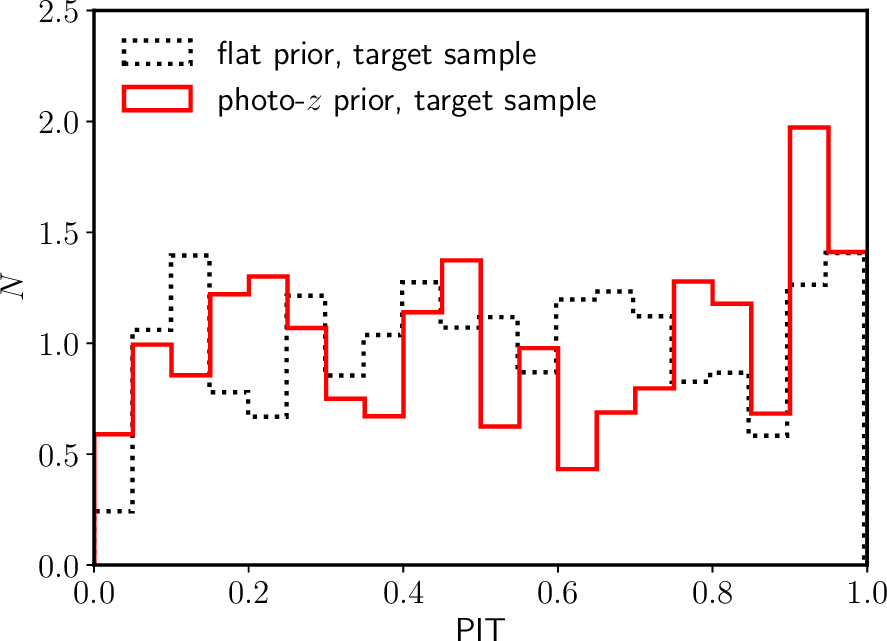} 
\end{tabular}
\caption{Same as Fig.~\ref{fig:PIT}, except that this refers to real data from the KiDS+VIKING-450 photometric survey and the VVDS-DEEP2 target sample. The sample is selected with a $\sigma_{z_{\rm p}}<0.6$ threshold in the \photoz{} uncertainties.}
\label{fig:PITKIDS}
\end{figure*}

\subsection{The KiDS+VIKING-450 survey}
\label{subsec:KIDS}

We now study a less favourable case, where the photometric survey has a similar depth as the target sample. We measure the mean redshift in five tomographic bins extracted from the KiDS+VIKING-450 imaging survey, which covers 341 deg$^2$ \citep{wright19}. The survey combines the $ugri$-band photometry from KiDS with $ZYJHK_\mathrm{S}$ bands from VISTA Kilo degree Infrared Galaxy (VIKING) photometry. We adopt the method described in Sect.~\ref{subsec:photoz} to measure the \photoz{}. This leads to a \photoz{} quality comparable to that obtained by \citet{wright19}, where $\sigma_{\rm NMAD}\sim 0.045$ at $z<0.9$ and $\sigma_{\rm NMAD}\sim0.079$ at $z>0.9$. Those \photoz{} are used to define five tomographic bins over the photometric redshift interval $0.1<z<1.2$, as in \citet[][]{hildebrandt20}.

The KiDS+VIKING-450 survey encompasses the VVDS \citep{lefevre05} and DEEP2 \citep{Newman:2013ha} fields, which contain spectroscopic redshifts. We aim at retrieving the mean redshift of the VVDS/DEEP2 galaxies. By only selecting galaxies with secure spectroscopic redshifts and counterparts in the KiDS+VIKING-450 catalogue, we build
a target sample of 5794 galaxies \footnote{We limit the risk of incorrect association between the photometric and spectroscopic sources by allowing a maximum angular separation of $0\overset{\arcsecond}{.}3$ in the match between KiDS-VINKING+450 and VVDS/DEEP2 catalogues.}. The DEEP2 sample has been selected at $R<24.1$ and $z>0.7$, while the VVDS sample is purely magnitude limited at $i<24$. Our target sample covers the full redshift range of interest $0.1<z<1.2$, with magnitude limits similar to those used for the KiDS+VIKING-450 cosmic shear analysis \citep{hildebrandt20}.

The KiDS+VIKING-450 imaging survey also covers the COSMOS field, and we use the existing \specz{} in the COSMOS field as the training sample. We note that the training and target samples are located in different fields. Therefore, the sample variance may impact our results. The COSMOS training sample contains $13\,817$ galaxies from the KiDS+VIKING-450 survey, after applying a redshift confidence selection. This highly heterogeneous sample combines various spectroscopic surveys covering a large range of magnitudes and redshifts \citep[see Sect.~\ref{subsec:specDescription} and ][for more details]{laigle16}.

We present our results in Table \ref{tabKIDS} for the five considered tomographic bins. 
The upper section of the table shows the fiducial case, where a $\sigma_{z_{\rm p}}<0.3$ \photoz{} uncertainty selection is applied. 
The direct calibration produces a bias of $|\Delta_{\avz}|<0.01\,(1+z)$, except in the lowest tomographic bin ($0.1<z<0.3$) where it reaches $|\Delta_{\avz}|=0.02\,(1+z)$. Using the debiased \zPDF{} method, we find $|\Delta_{\avz}|\,\lesssim\,0.01\,(1+z)$. In that case, the $\sigma_{z_{\rm p}}<0.3$ selection removes between 20\% and 44\% of the full KiDS+VIKING-450 sample\footnote{The representation fraction changes in each tomographic bin due to correlations between \specz{} and \photoz{} uncertainty.}. If we relax the selection on the \photoz{} error, as presented in the lower section of Table \ref{tabKIDS}, the bias $\Delta_{\avz}$ increases with the debiased \zPDF{} approach, as found in the simulation. Nonetheless, $\Delta_{\avz}$ remains around 1\%, which corresponds to an accuracy comparable to that obtain with direct calibration. We note that the \zPDF{} debiasing technique with the \photoz{} prior performs significantly better than with the flat prior. Figure \ref{fig:PITKIDS} illustrates the impact of the \photoz{} prior in recovering the shape of the redshift distribution, where we can see a clear improvement below the main mode (bottom left panel). This result is confirmed in the other tomographic bins.

The depth of the KiDS imaging survey is similar to the one we simulate for DES (5$\sigma$ sensitivity between 23.6 and 25.1), while the VIKING photometry is much shallower than the \Euclid{} one (between 21.2 and 22.7 for VIKING). It is therefore encouraging to find a bias similar to that expected from the simulation in the DES/\Euclid{} configuration, even with shallower imaging. We emphasise that our estimate is performed in the worst possible conditions: (1) our training sample does not cover the same colour/magnitude space as our target sample as shown in \citet{wright20}, (2) the photometric calibration could vary from field-to-field, and (3) some failures in the \specz{} target sample could bias the mean redshift considered as the truth. We know that a fraction of the target \specz{} could include catastrophic failures, possibly biasing our estimate of $\avz_{\rm true}$. Indeed, flag 3 in VVDS and DEEP2 are expected to be $97\%$ and $95\%$ correct, respectively, suggesting that a few percent of failures may be present in those samples, thereby introducing a bias in the true mean redshift $\avz_{\rm true}$ of more than 0.01, according to Fig.~\ref{fig:testErr}. The presence of such fraction of failures remains difficult to verify. A comparison between duplicated observations in DEEP2 shows that the fraction of failures should be at maximum $1.6\%$ \citep{Newman:2013ha}. 

Finally, we note that our various selections on $\sigma_{z_{\rm p}}$ prevent us from directly comparing the recovered redshift distributions with those published in \citet{wright19} and \citet{joudaki20}. Indeed, our selection $\sigma_{z_{\rm p}}$ preferentially removes the faintest galaxies from the sample, thus shifting the intrinsic redshift distribution towards lower redshifts than expected for the full KiDS+VIKING-450 sample. 

\begin{table*}
\begin{center}
\begin{tabular}{rrrrrrrrr} 
z$_{\rm min}$ & z$_{\rm max}$ & \% kept & N$_{\rm train}$ & \multicolumn{1}{c}{direct}  &  \multicolumn{1}{c}{\zPDF{} w/}    &  \multicolumn{3}{c}{\zPDF{} w/}  \\
              &               &         &                 &  \multicolumn{1}{c}{calib.}  
 & \multicolumn{1}{c}{flat prior}  & \multicolumn{3}{c}{photo-$z$ prior} \\
              &              &          &                 &  \multicolumn{1}{c}{$[10^{-2}]$} &   \multicolumn{1}{c}{$[10^{-2}]$} & \multicolumn{3}{c}{$[10^{-2}]$} \\
\hline \hline\\
\multicolumn{9}{c}{$\sigma_{z_{\rm p}}<0.3$} \\
\hline 

   0.10 &    0.30 &   79.80 &  1192.00 &     1.72  &  2.78  & &  0.94 & \\
   0.30 &    0.50 &   72.10 &  2156.00 &     0.64  &  0.33  & &  0.36 & \\
   0.50 &    0.70 &   55.60 &  1497.00 &    -0.57  & -0.88  & & -0.28 & \\
   0.70 &    0.90 &   68.70 &  1822.00 &    -0.65  & -1.38  & & -0.89 & \\
   0.90 &    1.20 &   62.00 &   892.00 &     0.10  &  0.29  & & -0.22 & \\
  
  \hline \\

\multicolumn{9}{c}{$\sigma_{z_{\rm p}}<0.6$}\\
\hline 
 
   0.10 &    0.30 &   96.60 &  1318.00 &     1.34  &     3.19  & &   -0.88  \\
   0.30 &    0.50 &   89.40 &  2321.00 &    -0.56  &     0.48  & &   -0.40  \\
   0.50 &    0.70 &   80.80 &  1845.00 &    -1.26  &    -2.60  & &   -1.50  \\
   0.70 &    0.90 &   89.60 &  2094.00 &    -0.34  &    -1.75  & &   -0.79  \\
   0.90 &    1.20 &   81.70 &  1057.00 &     0.38  &     1.16  & &   -0.03  \\

\hline   \\

\multicolumn{9}{c}{$\sigma_{z_{\rm p}}<1.2$}\\
\hline 

   0.10 &    0.30 &   97.80 &  1326.00 &     1.37  &     3.50  & &   -1.01  \\
   0.30 &    0.50 &   93.90 &  2357.00 &    -0.38  &     0.90  & &   -0.46  \\
   0.50 &    0.70 &   88.20 &  1886.00 &    -0.92  &    -2.42  & &   -1.63  \\
   0.70 &    0.90 &   93.70 &  2131.00 &    -0.11  &    -1.67  & &   -0.92  \\
   0.90 &    1.20 &   90.40 &  1116.00 &     1.66  &     2.67  & &    0.43  \\

\hline  
  \end{tabular}
\caption{Differences between the mean redshifts reconstructed with different methods (direct calibration and debiased \zPDF{}) and $\avz_{\rm true}$, divided by $(1+\avz_{\rm true})$. The KiDS+VIKING-450 survey is split in five tomographic bins. We use VVDS/DEEP2 as target sample, and COSMOS as the training one. In the top part of the table, \photoz{} are selected with $\sigma_{z_{\rm p}}<0.3$, while the bottom parts show a selection at $\sigma_{z_{\rm p}}<0.6$ and $\sigma_{z_{\rm p}}<1.2$. The fraction of galaxies kept after this selection is also shown (`\% kept'). We apply the same definition as \citet{wright20} to define the loss of photometric sources (their Eq.~1), including shear weights.}
\label{tabKIDS}
\end{center}
\end{table*}

\section{Summary and conclusion}
\label{sec:summary}

This paper investigates the possibility of measuring the mean redshift $\avz$ of a target sample of galaxies, in ten tomographic bins from $z=0.2$ to $z=2.2$,  with an accuracy of $|\Delta_{\avz}|\,<\,0.002\,(1+z)$, as stipulated by the \Euclid{} mission requirements on cosmic shear analysis. Naturally, the conclusions presented here are equally applicable to all current and future surveys where redshift calibration is a relevant challenge.

We apply two approaches which are foreseen for the \Euclid{} mission : a direct calibration of $\avz$ with a spectroscopic training sample and the combination of individual \zPDF{} to reconstruct the underlying redshift distribution. This paper analyses in detail several factors which could impact these approaches and provide recommendations to make them successful.

We use the Horizon-AGN hydrodynamical simulation \citep{dubois14}, which allows a large diversity of modelled SED, and create 18 \Euclid{}-like mock catalogues, with different realisations of the photometric noise. We simulate two possible configurations, which should encompass the range of sensitivities of future imaging available for \Euclid{}: (1) a shallow configuration combining DES and \Euclid{}, and (2) a deep configuration combining LSST and \Euclid{}. We measure the \photoz{} of the simulated galaxies using the template-fitting code \lephare{}, as performed in \citet{laigle19}. Such procedure produces photometric redshifts with complex zPDF{}, realistic biases, and catastrophic failures. We also assume different characteristics for the spectroscopic training samples associated to the mock catalogues. We consider several selection functions, several sample sizes, and include possible failures in the \specz{}.

We first test the direct calibration approach, where the redshift distribution is directly estimated from existing spectroscopic redshifts in a training sample, applying necessary weights to match this distribution to the target sample. We find that this approach is efficient in recovering the mean redshift with an accuracy of $0.002\,(1+z)$. The method is successful when based on a representative spectroscopic coverage (uniform or SOM), but the weighting scheme is not sufficient to correct for the heterogeneity in the COSMOS-like training sample at the level required by \Euclid{}. This method is stable and robust, and does not require deep photometry such as that from LSST. However, we find that the recovered mean redshift is extremely sensitive to the presence of catastrophic failures in spectroscopic redshift measurement. To recover unbiased estimates of $\avz$, a careful quality assessment of the spectroscopic redshifts must guarantee a fraction of failures below 0.2\%. 

We then investigate the possibility of reconstructing the redshift distribution from the \zPDF{} produced by a template-fitting \photoz{} code. As expected, we find that the quality of the initial \zPDF{} is not sufficient to measure $\avz$ with an accuracy better than $|\Delta_{\avz}|\,<\,0.01$. 
We test the method by \citet{bordoloi10} to debias the \zPDF{}. We improve it by taking into account an appropriate prior, combined with an iterative correction of the \zPDF{}. Our results are summarised below.
\begin{itemize}
    \item The mean redshift accuracy inferred from the debiased \zPDF{} is systematically improved when compared to the one inferred from the initial \zPDF{} (by up to a factor ten). 
    \item This method is weakly sensitive to the fraction of \specz{} failures.
    \item Imaging depth is the primary factor in determining the effectiveness of the debiasing technique. We reach the \Euclid{} requirement when combining \Euclid{} and LSST ground-based images. 
    \item Insufficient imaging depth can be compensated by selecting well peaked \zPDF{}, but introduces considerable losses to the target sample number density. A balance should therefore be established between the accuracy of  $\avz{}$ and the statistical signal of the cosmic shear analysis.
\end{itemize}

We test the two approaches on real data sets from COSMOS and KiDS+VIKING-450, and confirm that a high signal-to-noise in the photometry is essential for an accurate estimate of $\avz$ using the debiased \zPDF{} approach. In the less favourable case (KiDS+VIKING-450), where the photometric sample and a \specz{} target sample are approximately of equal depth, we reach an accuracy around $0.01\,(1+z)$ on $\avz$, as expected from the simulation and other works \citep[e.g.][]{wright20}. We confirm the trends observed in the simulation and find that including the prior in  the debiasing technique produces significantly better results.

We conclude that both methods could foreseeably provide independent and accurate inferences of tomographic bin mean redshifts for \Euclid{}. We find that the current \Euclid{} baseline to measure $\avz$ with a direct calibration approach and a SOM training sample is robust with respect to the imaging survey depth. However, we recommend that training samples, such as C3R2 \citep{masters19}, insure a purity level above $99.8\%$. We also find that the sum of the debiased \zPDF{} could be sufficient to measure $\avz{}$ at the \Euclid{} requirement, with currently ongoing spectroscopic surveys. However, we recommend this method only in areas covered with deep optical data. The two methods should be applied simultaneously with current planning of the \Euclid{ } survey and provide complementary and independent estimate of $\avz{}$.

Finally, our work still suffers from several limitations that we still need to investigate. We neglect the catastrophic failures within the \photoz{} sample created by misclassified stars or AGN, or by the galaxy blending. A residual contamination of these populations in the tomographic bins could affect both approaches to redshift calibration. Moreover, we do not consider sample variance effects, since the Horizon-AGN simulation covers only 1 deg$^2$. We would benefit from a larger simulated area to test the impact of sample variance. Nonetheless, our results here present a largely positive outlook for the challenge of tomographic redshift calibration within \Euclid{}.   

\begin{acknowledgements}
We thank the OU-PHZ of \Euclid{} for all the useful discussions along these years.
OI acknowledges the funding of the French Agence Nationale de la Recherche for the project `SAGACE'. NM acknowledges support from a CNES fellowship. H. Hildebrandt is supported by a Heisenberg grant of the Deutsche Forschungsgemeinschaft (Hi 1495/5-1) as well as an ERC Consolidator Grant (No. 770935). A.H. Wright is supported by the ERC Consolidator Grant (No. 770935). This work  relied on the HPC resources of CINES (Jade) under the allocation 2013047012 and c2014047012 made by GENCI and on the Horizon Cluster hosted by Institut d'Astrophysique de Paris. ID acknowledges that he received funding from the European Union’s Horizon 2020 research and innovation programme under the Marie Sk\l{}odowska-Curie grant agreement No. 896225. We  warmly thank S.~Rouberol for running  the  cluster on which the simulation was  post-processed. This research  is also partly supported by the Centre National d'Etudes Spatiales (CNES). We would also like to recognise the contributions from all of the members of the COSMOS team who helped in obtaining and reducing the large amount of multi-wavelength and spectroscopic data. Based on observations made with ESO Telescopes at the La Silla Paranal Observatory under programme IDs 177.A-3016, 177.A-3017, 177.A-3018 and 179.A-2004, and on data products produced by the KiDS consortium. The KiDS production team acknowledges support from: Deutsche Forschungsgemeinschaft, ERC, NOVA and NWO-M grants; Target; the University of Padova, and the University Federico II (Naples). SA thank the support PRIN MIUR2015 "Cosmology and Fundamental Physics: Illuminating the Dark Universe with Euclid".\AckEC
\end{acknowledgements}

\bibliographystyle{aa}
\bibliography{paper.bbl}

\begin{appendix}

\section{Idealised test of the debiasing procedure}\label{idealized}

We present in this appendix how we generated a simplified mock catalogue, in comparison to the one presented in Sect.~\ref{sec:simulation}. We still use the Horizon-AGN mock catalogue. Rather than using the \photoz{} produced by \lephare{}, however, we generate an idealised \photoz{}. We randomise the true redshift assuming a Gaussian distribution with $\sigma=\sigma_{\rm true}$, where $\sigma_{\rm true}$ is defined as the median value of the \lephare{} \photoz{} errors. We then bias these \photoz{} by applying a systematic shift of $\Delta_{z_{\rm p}}=-0.05$. We associate with each galaxy a likelihood defined as:
\begin{equation}
{\mathcal L }(z) = \frac{1}{A\, \sigma_{\rm true} \sqrt{2\pi}} \exp \left[ -\frac{1}{2} \left( \frac{z-z_{\rm p}}{A\, \sigma_{\rm true}} \right)^2 \right],
\end{equation}
where the factor $A$ allows us to mimic an under-estimation (over-estimation) of the \photoz{} uncertainties, if $A<1$ ($A>1$).
In this way we can check, using a simplified simulation, if we are able to recover the true mean redshift despite having a bias in the \photoz{} and their associated likelihood.

\begin{figure*}
\begin{tabular}{l l}
\includegraphics[width=8.5cm]{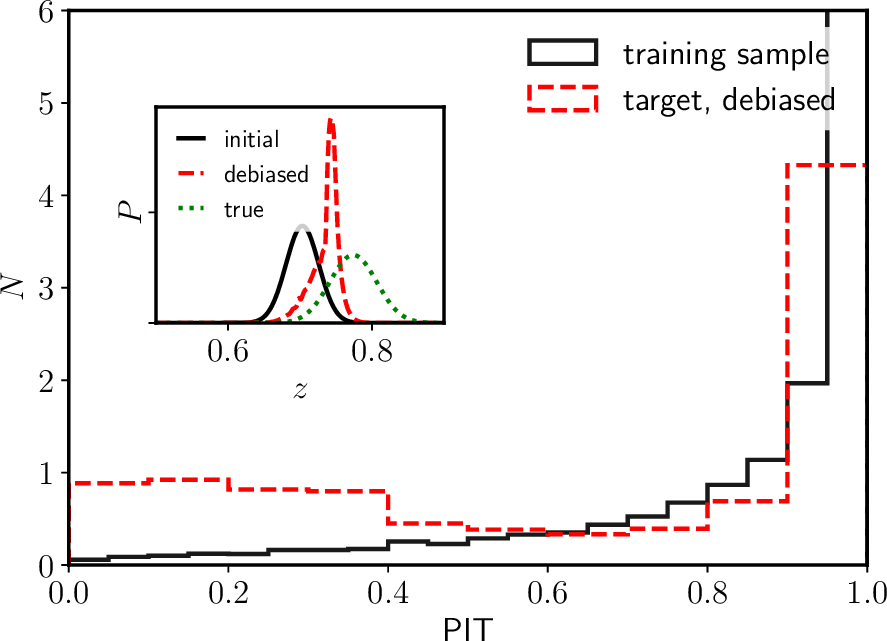}  &
\includegraphics[width=8.5cm]{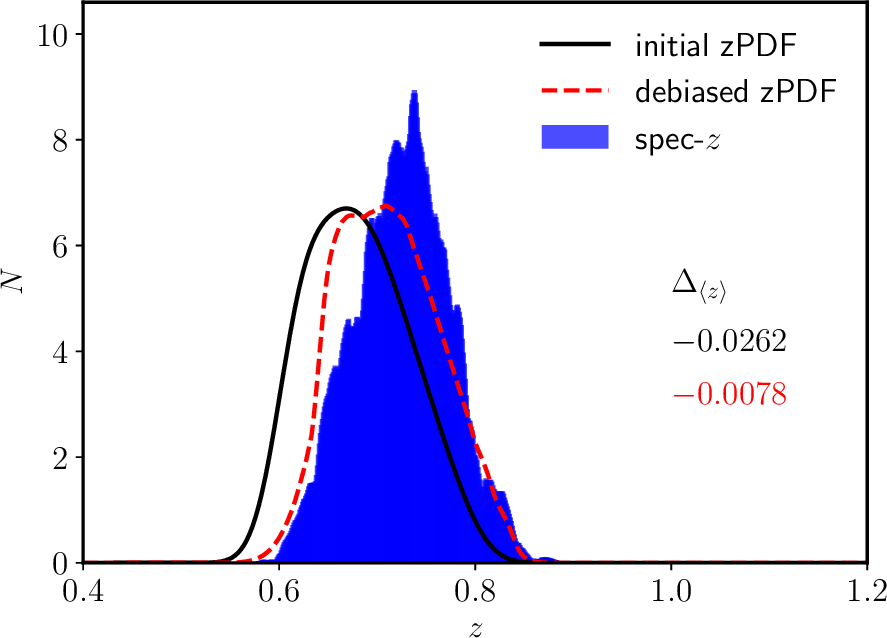} \\   
\includegraphics[width=8.5cm]{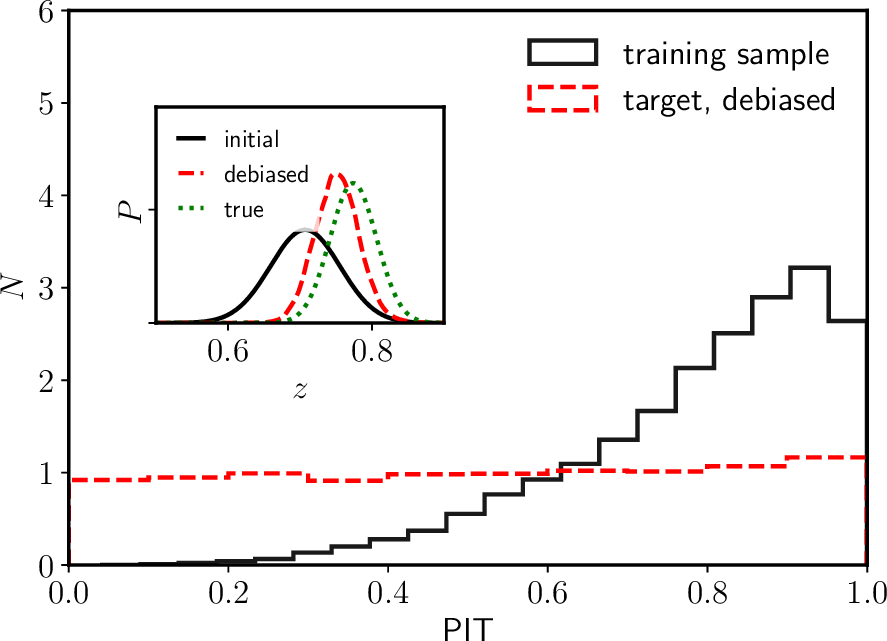}  &
\includegraphics[width=8.5cm]{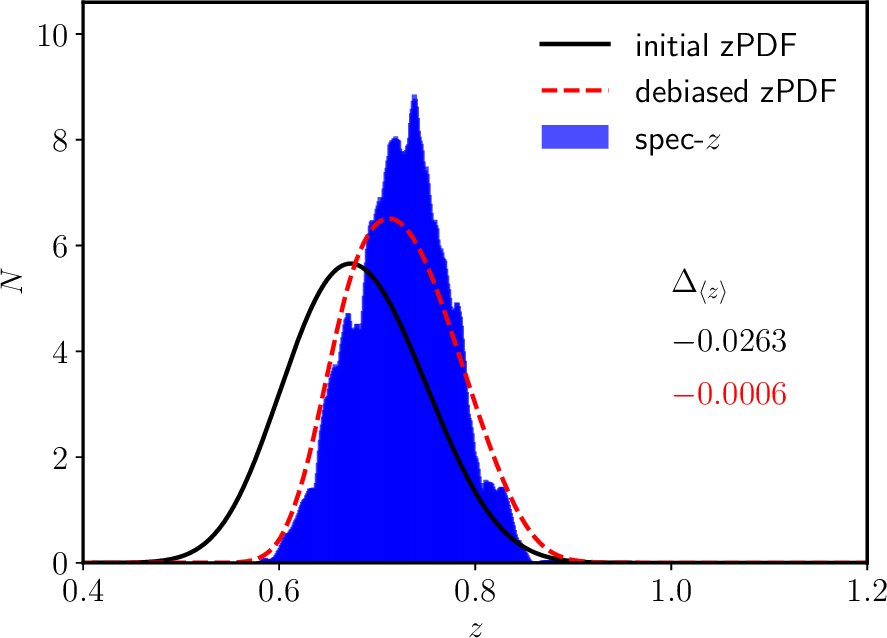} 
\end{tabular}
\caption{Example of PIT distribution (left) and redshift distribution (right) for a tomographic bin selected at $0.6 < z_{\rm p} < 0.8$. The top and bottom panels assume \photoz{} errors that are under-estimated ($A=0.7$) and over-estimated ($A=1.5$), respectively. The PIT distribution used to correct the \zPDF{} is shown with the solid black line. The inset shows an example of the debiased \zPDF{} for one galaxy (selected randomly). The resulting PIT distribution, after debiasing, is shown in dashed red. The true $N(z)$ is shown with the blue histogram in the right panels. The $N(z)$ reconstructed using the initial and the debiased \zPDF{} are shown with black solid lines and red dashed lines, respectively.
\label{fig:simple}}
\end{figure*}

We apply the same method as described in Sect.~\ref{subsec:BordoMethod} to recover the mean redshift, assuming a flat prior. We select galaxies in a tomographic bin at $0.6<z_{\rm p}<0.8$. Two examples are given in Fig.~\ref{fig:simple}. The top (bottom) panels assume $A=0.7$ ($A=1.5$), i.e. \photoz{} errors that are under-estimated (over-estimated).

We find that as long as $A>1$, the method is efficient in recovering the mean redshift. However, if the original \zPDF{} are too narrow ($A<1$), the final correction is unstable. We find the same result by testing several values of $A$ and several values of the bias. Therefore, we conclude that \photoz{} errors should be preferentially  overestimated in the application of the debiased \zPDF{} method.

As a result, when applying our template-fitting code to the Horizon-AGN simulated galaxies, we  simply multiply the flux uncertainties by a constant factor to ensure that we are working in this regime. Specifically, for comparison to the \photoz{} measured by \citet{laigle19}, we multiply the flux uncertainties by a factor 1.5 and impose a minimal error of $\Delta m = 0.01$ in each band.  

\section{Mean redshift of the LEGA-C survey in COSMOS}
\label{appendix:COSMOS-LEGAC}

The goal in this section is to retrieve the mean redshift of the LEGA-C galaxies \citep{vanderwelt16} selected in the tomographic bin $0.7<z_{\rm p}<0.9$. We base our estimate of $\avz{}$ on the COSMOS photometry and associated \specz{} (excluding LEGA-C \specz{} from the training). Then, we compare the estimated mean redshift with the true one (known from LEGA-C \specz{}). In such configuration, the photometry is much deeper than the selection limit of the target sample.

\noindent {\bf The COSMOS photometry.} We use the photometric catalogue from \citet{laigle16}, but keeping only the ten broad bands: $u$, $B$, $V$, $r$, $i$, $z$, $Y$, $J$, $H$, $K$. We adopt exactly the same method to compute the \photoz{} as the one described in Sect.~\ref{subsec:photoz}. As described in Sect.~\ref{subsec:BordoMethod}, we inflate our photometric flux uncertainties within the input photometric catalogue by a factor of two, to allow for better debiasing. 

\noindent {\bf LEGA-C target sample.} We select a spectroscopic sample as robust as possible, to ensure that the uncertainty on the mean redshift of the target sample (considered as the truth) is known with an accuracy better than 0.002. The LEGA-C spectroscopic survey in the COSMOS field provides such a target sample. This spectroscopic sample is built using the high-resolution ($R=3000$) mode of VIMOS spectrograph, targeting galaxies at $0.6<z<1$ selected in the $K_{\rm s}$-band to have a stellar mass $M_\star>10^{10}\Msol$. Given the resolution and the SN reached by the LEGA-C spectra (with 20\,h of exposure per spectrum), and the numerous absorption/emission lines detected, we can safely assume that this sample does not include any catastrophic spectroscopic failures. We match the LEGA-C Data Release 2 galaxies \citep{straatman19} to the COSMOS2015 catalogue on-sky, allowing a maximum angular separation of $0\overset{\arcsecond}{.}2$ in the association. This reduces the risk of incorrectly associating spectra to our COSMOS2015 photometry. Our LEGA-C target sample thus contains 1213 galaxies, with a median $i$-band magnitude of 21.45.

\noindent  {\bf The COSMOS training sample.} Since the constraint in terms of completeness and purity is less stringent for the training sample, we randomly choose 50\% of all the \specz{} available in COSMOS, irrespective of  magnitude. We remove all the LEGA-C sources from the training sample, and combine the \specz{} from multiple surveys, namely: zCOSMOS-Bright and Faint \citep{lilly07}, FMOS \citep{kashino19}, and C3R2 \citep{masters19}. We select only spectra with either `high confidence' or `certain' redshift confidence flags \citep[corresponding to flag 3--4 in the VVDS redshift confidence flagging system of][]{lefevre05}, in order to select only the most reliable redshifts for our training set. Still, the magnitude and colour distributions differ between the training and the target sample. We thus apply a weight to each galaxy of the training sample to reproduce the global properties of the target sample. Those weights are derived by projecting the target sample over the SOM, as described in Sect.~\ref{sec:direct} for the COSMOS-like sample. We construct our SOM here using magnitudes, colours, and the \photoz{} associated with the training sample sources. We adopt a $10\times10$ SOM, smaller than the one used in Horizon-AGN, because of the limited size of the target sample.

\noindent  {\bf Application.} We select all sources with \photoz{} in the range $0.7<z_{\rm p}<0.9$ (we choose this redshift range since it needs to overlapp with LEGA-C). We create 300 realisations with a random selection of the training sources. The target sample consists of $493$ galaxies, of which around $5$\%  have $\sigma_{z_{\rm p}}>0.3$ and are subsequently removed. We estimate the mean redshift of the target sample using the direct calibration, direct \zPDF{}, and debiased \zPDF{} approaches, and compare these with the true $\avz$ of the target sample. For the direct calibration approach, we obtain a bias of $\biaszmeanB{}=0.00032$ and a scatter of $\sigma_{\Delta z}=0.00135$; an accuracy well within the \Euclid{} requirement. Secondly, we estimate $\avz$ using the initial \zPDF{} without debiasing. We obtain a mean redshift biased by $\biaszmeanB{}\,>\,-0.013$, which is six times larger than the \Euclid{} requirement. Finally, we debias the \zPDF{} using the PIT distribution as discussed in Sect.~\ref{subsec:BordoMethod}. In that case, we obtain a mean redshift with a bias of $\biaszmeanB{}=-0.00046$ ($\biaszmeanB{}=-0.00008$) and a scatter of $\sigma_{\Delta z}=0.00073$ ($\sigma_{\Delta z}=0.00074$) assuming the \photoz{} (flat) prior. Therefore, in the case of a target sample associated with much deeper photometry, we reach the $0.002\,(1+z)$ accuracy requirement of \Euclid{}, either using the direct calibration or debiased \zPDF{} approaches.

\end{appendix}

\end{document}